\documentclass[12pt,twoside]{article}
\usepackage{mathpazo} 
\usepackage[mathscr]{eucal}
\usepackage{amsmath,amsfonts,amssymb,amsthm}
\usepackage{times}
\usepackage{hyperref}

\def\d_Vphi{\text{d}_V\hspace{-0.06em}\phi}
\def\d_Vphibar{\text{d}_V\hspace{-0.06em}\bar\phi}
\def\d_Vxi{\text{d}_V\hspace{-0.06em}\xi}

\def\be{\begin{eqnarray}}
\def\ee{\end{eqnarray}}
\def\beann{\begin{eqnarray*}}
\def\eeann{\end{eqnarray*}}
\def\beq{\begin{equation}}
\def\eeq{\end{equation}}
\def\ba{\begin{array}}
\def\ea{\end{array}}
\def\ben{\begin{enumerate}}
\def\een{\end{enumerate}}
\def\bea{\begin{eqnarray}}
\def\eea{\end{eqnarray}}

\def\5{\bar }
\def\6{\partial }
\def\7{\hat }
\def\4{\tilde }
\advance\textheight\topskip
\renewcommand{\tilde}{\widetilde}
\renewcommand{\hat}{\widehat}

\renewcommand{\simeq}{\cong}

\newcommand{\bref}[1]{\textbf{\ref{#1}}}

\newcommand{\dd}{\partial}
\renewcommand{\d}{\partial}

\newcommand{\binner}[2]{%
  {\langle}\kern-4.15pt{\langle}#1{,}\,#2{\rangle}\kern-4.15pt{\rangle}}

\newcommand{\half}{\frac{1}{2}}

\newcommand{\ffrac}[2]{\raisebox{.5pt}%
  {\footnotesize$\displaystyle\frac{#1}{#2}$}\kern1pt}

\newcommand{\ddl}[2]{\ffrac{\dd #1}{\dd #2}}

\def\cH{\mathcal{H}}

\def\cL{\mathcal{L}}
\def\cM{\mathcal{M}}

\def\cP{\mathcal{P}}

\numberwithin{equation}{section} \makeatletter
\@addtoreset{equation}{section}

\DeclareFontFamily{OT1}{rsfs}{} \DeclareFontShape{OT1}{rsfs}{m}{n}{
<-7> rsfs5 <7-10> rsfs7 <10-> rsfs10}{}
\DeclareMathAlphabet{\mycal}{OT1}{rsfs}{m}{n}
\def\scri{{\mycal I}}%

\begin{document}

\def\mytitle{Dual dynamics of three dimensional asymptotically flat
  Einstein gravity at null infinity}

\pagestyle{myheadings} \markboth{\textsc{\small G.~Barnich, H.~Gonz\'alez}}{%
  \textsc{\small Dual dynamics of 3d flat gravity at $\scri$}}
\addtolength{\headsep}{4pt}

\begin{flushright}\small
ULB-TH/12-24\end{flushright}

\begin{centering}

  \vspace{1cm}

  \textbf{\Large{\mytitle}}


  \vspace{1.5cm}

  {\large Glenn Barnich$^{a}$}

\vspace{.5cm}

\begin{minipage}{.9\textwidth}\small \it  \begin{center}
   Physique Th\'eorique et Math\'ematique \\ Universit\'e Libre de
   Bruxelles and International Solvay Institutes \\ Campus
   Plaine C.P. 231, B-1050 Bruxelles, Belgium
 \end{center}
\end{minipage}

\vspace{.5cm}

{\large Hern\'an A. Gonz\'alez$^{b}$}

\vspace{.5cm}

\begin{minipage}{.9\textwidth}\small \it \begin{center}
   Departamento de F\'{\i}sica, P. Universidad Cat\'olica de
Chile \\ Casilla 306, Santiago 22, Chile
\end{center}
\end{minipage}

\end{centering}

\vspace{1cm}

\begin{center}
  \begin{minipage}{.9\textwidth}
    \textsc{Abstract}. Starting from the Chern-Simons formulation, the
    two-dimensional dual theory for three-dimensional asymptotically
    flat Einstein gravity at null infinity is constructed. Solving the
    constraints together with suitable gauge fixing conditions gives
    in a first stage a chiral Wess-Zumino-Witten like model based on the
    Poincar\'e algebra in three dimensions. The next stage involves a
    Hamiltonian reduction to a BMS3 invariant Liouville
    theory. These results are connected to those originally derived in
    the anti-de Sitter case by rephrasing the latter in a suitable
    gauge before taking their flat-space limit.
  \end{minipage}
\end{center}

\vfill

\noindent
\mbox{}
\raisebox{-3\baselineskip}{%
  \parbox{\textwidth}{\mbox{}\hrulefill\\[-4pt]}}
{\scriptsize$^a$Research Director of the Fund for Scientific
  Research-FNRS Belgium. E-mail: gbarnich@ulb.ac.be\\
$^b$ E-mail: hdgonzal@uc.cl }

\thispagestyle{empty}
\newpage

\begin{small}
{\addtolength{\parskip}{-1.5pt}
 \tableofcontents}
\end{small}
\newpage

\section{Introduction}
\label{sec:introduction}

In the context of holographic approaches
\cite{tHooft:1993gx,Susskind:1994vu,Maldacena:1998re} to gravitational
theories, $2+1$ dimensional models play a prominent role because there
is detailed quantitative understanding both of the bulk theories and
of their two dimensional dual.

For the particular case of three dimensional gravity, the Chern-Simons
formulation \cite{Achucarro:1986vz,Witten:1988hc} can be used to good
effect. Indeed, the dual theory on closed spatial sections is obtained
simply by solving the constraints inside the Chern-Simons action
\cite{Witten:1988hf,Moore:1989yh,Elitzur:1989nr} giving rise to a
(chiral) Wess-Zumino-Witten model \cite{Witten:1983ar}. Unlike in
most conformal field theory considerations, the relevant groups in
applications to gravity are non-compact or non semi-simple, that is
$SO(2,2)$ in the AdS case and $ISO(2,1)$ in the flat
case. Furthermore, the spatial section is a plane and the choice of
boundary conditions plays a crucial role in determining the dual
theory.

In the AdS case, these questions have been addressed in
\cite{Coussaert:1995zp} (see also
\cite{Banados:1994tn,Carlip:1994gy,Carlip:1998uc,Banados:1998ta,%
  Carlip:1998qw,1999AIPC..484..147B} for related considerations). In
particular, the chiral decomposition $\mathfrak{so}(2,2)\simeq
\mathfrak{sl}(2,\mathbb R)\oplus \mathfrak{sl}(2,\mathbb R)$ allows
one to apply standard techniques for semi-simple algebras in each
sector. The first stage of the reduction then involves a formulation
in terms of decoupled chiral models
\cite{Floreanini:1987as,Bernstein:1988zd,Sonnenschein:1988ug,%
  Salomonson:1988mk} that combine into a standard Wess-Zumino-Witten
model in a well-understood way (see
e.g.~\cite{Stone:1989vg,Chu:1991pn,Caneschi:1996sr}). In a second
stage, the gravitational boundary conditions allow for a further
simplification by implementing a standard Hamiltonian reduction from
the $SL(2,\mathbb{R})$ Wess-Zumino model to Liouville theory
\cite{Forgacs:1989ac,Alekseev1989719,Bershadsky:1989mf}.

The main purpose of the present paper is to construct the dual theory
for three dimensional asymptotically flat gravity at null infinity and
to establish its connection with the AdS results. Apart from shedding
light on details of holography in backgrounds that are not AdS, such a
dual theory is liable to play a role as a toy model for cosmological
scenarios (see e.g.~\cite{Cornalba:2003kd} and references therein) due
to the existence of time-dependent cosmological solutions in this
context \cite{Barnich:2012aw}.

Not surprisingly, a detailed analysis of the Chern-Simons to
Wess-Zumino-Witten reduction for the Poincar\'e algebra
$\mathfrak{iso}(2,1)$ does exist \cite{Salomonson:1989fw}. We will
however have to adjust the analysis to the case at hand. Indeed, for
our purpose, it will be more convenient to work with the spinor rather
than the vector representation of $\mathfrak{so}(2,1)$ in order to
connect AdS and flat space results. Furthermore, the boundary
conditions that have been used are not directly related to those of
asymptotically flat spacetimes at null infinity. Implementing the
appropriate boundary conditions modifies the resulting chiral
Wess-Zumino-Witten model and is important in order to have as rich a
dynamics in the flat as in the AdS case
\cite{Barnich:2006avcorr,Barnich:2010eb} with a direct connection
between the two asymptotic regimes \cite{Barnich:2012aw} (see also
\cite{Bagchi:2012cy}). In turn, this is crucial in order to repeat the
semi-classical arguments for a microscopic explanation of the BTZ
black hole entropy \cite{Strominger:1998eq} of the corresponding
asymptotically flat cosmological solutions
\cite{Barnich:2012xq,Bagchi:2012he}.

As for other non semi-simple algebras (see e.g.~\cite{Nappi:1993ie}),
the chiral Wess-Zumino-Witten like model for $\mathfrak{iso}(2,1)$
admits a globally well-defined two-dimensional action. The central
extension in the associated current algebra affects the brackets
between rotation and translations generators. In this case, the
Hamiltonian reduction of the model gives rise to a BMS3 invariant
Liouville type theory that is discussed in more detail in the
companion paper \cite{Barnich:2012rz}.

The paper is organized as follows. Instead of using asymptotic
conditions, we consider instead a suitable gauge fixed form of the
metric. This is not the more standard Fefferman-Graham form in the AdS
case, but rather a BMS type gauge that allows for a parallel treatment
of both the AdS and the flat case. After quickly reviewing the general
solution to the three-dimensional Einstein's equations, we provide in
section \bref{sec:bms-gauge-boundary} explicit expressions for the
associated dreibeins and spin connections. 

Section \bref{sec:group-elements} is devoted to constructing the
associated group elements. The field corresponding to the Cartan
generator of $\mathfrak{sl}(2,\mathbb{R})$ can then be related to a
standard Liouville field in the AdS case and a BMS Liouville field in
the flat case. In particular, the overall normalization that has been
left unspecified in \cite{Barnich:2012rz} can be fixed at this
stage. The limit relating the group elements of the AdS to the flat
case is provided and shows how the explicit time dependence emerges
from this point of view.

The remainder of the work consists in deriving the equations for the
group elements on the level of action principles.  In a first step in
section \bref{sec:reduct-wess-zumino}, suitable boundary terms are
added to the Chern-Simons action in order to make the variational
problem well-defined for the gravitational solutions that we are
interested in. In terms of vielbeins and spin connections, this step
can be done in parallel for both AdS and flat space with an obvious
limit.

In Section \bref{sec:reduct-wess-zumino-1} and the associated
appendix, we first briefly recall results on the Chern-Simons to WZW
reduction for the AdS case, in particular how the reduction gives rise
in a first step to chiral $\mathfrak{sl}(2,\mathbb R)$ WZW models. We
then review the structure of these models from the point of view of
constrained Hamiltonian systems, including their current algebras and
classical conformal invariance. These steps can then be directly
generalized to the flat case, where an appropriate chiral
$\mathfrak{iso}(2,1)$ WZW model is constructed. Its general solution
involves a linear time-dependence and the $\mathfrak{iso}(2,1)$
current algebra is constructed in terms of Dirac brackets. BMS3
invariance of the model is established in terms of the current algebra
along standard lines. Finally, we show how to obtain the chiral
$\mathfrak{iso}(2,1)$ WZW like model as a flat limit limit of two
chiral $\mathfrak{sl}(2,\mathbb R)$ models.

In the last section~\bref{sec:Liouville}, the Hamiltonian reduction is
implemented. In the AdS case, they reduce the chiral models
$\mathfrak{sl}(2,\mathbb R)$ WZW models to free chiral bosons that
combine into Liouville theory in a standard way. In the flat case, a
free first order action principle is obtained that is related to the
BMS3 Liouville theory in a similar way.

In order to emphasize novel aspects, conventions, notations and
intermediate formulae that are relevant only to follow the details of
the computations are mostly relegated to the appendix.

In all the analysis, we have concentrated for simplicity on the boundary at
future null infinity. This corresponds to analysing Chern-Simons
theory with a spatial section that is a disk. In a more complete
analysis, other boundaries, sources in the interior and holonomies can
and should be taken into account by following the arguments in
\cite{Elitzur:1989nr,Henneaux:1999ib,Rooman:2000zi}.

Obvious generalizations of the present work consist in including in
the starting point Chern-Simons formulation the exotic term, i.e., the
Chern-Simons terms for the spin-connection \cite{Witten:1988hc}. The
inclusion of this term can be entirely captured through an extension
of the invariant metric that does not affect equations of motion or
constraints, but suitably modifies the current algebras. A related
generalization consists in repeating the analysis for topologically
massive gravity \cite{Deser:1981wh,Deser:1982vy}.

We have limited ourselves to the classical theory, but it should
obviously be interesting to consider quantum aspects of the
$\mathfrak{iso}(2,1)$ chiral Wess-Zumino theory and investigate for
instance to what extent the general analysis of
\cite{FigueroaOFarrill:1994hx,FigueroaOFarrill:1995cz} applies. These
questions will be addressed elsewhere.

\section{BMS gauge, fall-off conditions and general solution}
\label{sec:bms-gauge-boundary}

The BMS gauge consists in using the diffeomorphisms to put the metric in
the form 
\begin{equation}
  \label{eq:ojo}
   ds^2=e^{2\beta}\frac{V}{r} du^2-2e^{2\beta}
   dudr+r^2(d\phi-Udu)^2,
\end{equation}
in terms of three arbitrary functions $\beta, V, U$. Here $r$ is a
radial coordinate restricted to $r\in [\bar r,\infty)$, $u$ is a null
coordinate, while $\phi\in [0,2\pi]$ is an angular coordinate.
Associated dreibeins $e^a$ such that $ds^2=2e^0e^1+(e^2)^2$ can
be chosen as
\begin{equation}
  \label{eq:9}
  e^0=\half (e^{2\beta} \frac{V}{r}+r^2 U^2)du-e^{2\beta}
  dr-r^2 Ud\phi,\quad e^1=du,\quad 
  e^2=rd\phi.
\end{equation}

When imposing the fall-off conditions $\beta=o(1)=U$, the Einstein
equations can be solved exactly. They imply in particular the stronger
fall-off conditions 
\begin{equation}
  \label{eq:3}
  \frac{V}{r}=-\frac{r^2}{l^2}+O(1),\quad \beta=O(r^{-1}),\quad
  U=O(r^{-2}), 
\end{equation}
that can be used to complete the definitions of asymptotically
anti-de Sitter or flat space-times in BMS gauge. In the flat case, the
limit $l\to\infty$ is understood so that $\frac{V}{r}=O(1)$.

The exact solution is given by 
\begin{equation}
  \label{eq:8}
  ds^2=\left(-\frac{r^2}{l^2}+\mathcal{M}\right)du^2-2du
  dr+2\mathcal{N}du d\phi+r^2d\phi^2,
\end{equation}
where $\d_r\mathcal M=0=\d_r\mathcal N$ and 
\begin{equation}
  \label{eq:14}
  \d_u \mathcal M=\frac{2}{l^2}\d_\phi\mathcal N,\quad 2\d_u\mathcal
  N=\d_\phi\mathcal M. 
\end{equation}
The general solution to these equations is 
\begin{equation}
  \label{eq:10}
  \mathcal{M}=2(\Xi_{++}+\Xi_{--}),\quad 
\mathcal{N}=l(\Xi_{++}-\Xi_{--}),
\end{equation}
with $\Xi_{\pm\pm}=\Xi_{\pm\pm}(x^\pm)$, in the ${\rm AdS}$ case and
\begin{equation}
  \label{eq:11}
  \mathcal{M}=\Theta,\quad
  \mathcal{N}=\Xi+\frac{u}{2}\d_\phi \Theta,
\end{equation}
with $\Theta=\Theta(\phi)$ and $\Xi=\Xi(\phi)$ in the flat case. In
terms of the arbitrary functions, the conserved charges in the AdS
case associated to
$\xi=Y^+\d_++Y^-\d_-$,  $Y^\pm=Y^\pm(x^\pm)$, are given by
\begin{equation}
  \label{eq:109}
  Q_{Y^\pm}=\frac{l}{8\pi G}\int ^{2\pi}_0d\phi (Y^+\Xi_{++}+Y^-\Xi_{--}),
\end{equation}
if normalized with respect to the $M=0=J$ BTZ black
hole. In the flat case, they are associated to $\xi=(T+u Y')\d_u +
Y\d_\phi$, $T=T(\phi),Y=Y(\phi)$ and given by 
\begin{equation}
  \label{eq:110}
  Q_{T,Y}=\frac{1}{16 \pi G}\int^{2\pi}_0 d\phi (T \Theta+2Y \Xi),
\end{equation}
when normalized with respect to the null orbifold.

In the first order formalism, the equations of motion are
\begin{equation}
d\omega+\omega^2 +\frac{1}{l^2}e^2=0,\quad
d e+\omega e + e \omega =0. \label{eq:13}
\end{equation}
Associated dreibeins and spin connections are given by
\begin{equation}
  \label{eq:12}
\begin{split}
&  e^0=-\half(\frac{r^2}{l^2}-
  \mathcal{M})du-dr+\mathcal{N}d\phi\quad e^1=du,\quad e^2=rd\phi,\\
& \omega^0=\frac{\mathcal{N}}{l^2}du-\half(\frac{r^2}{l^2}-\mathcal
  M)d\phi,\quad\omega^1=d\phi,
\quad \omega^2=\frac{r}{l^2}du. 
\end{split}
\end{equation}
In particular, we note that for the gravitational solutions that we are interested
in 
\begin{equation}
  \label{eq:18}
  \omega^a_\phi=e^a_u,\quad
  \omega^a_u=\frac{1}{l^2} e^a_\phi,\quad
  \omega^a_r=0,\quad \delta e^a_r=0=\d_\phi e^a_r=\d_ue^a_r. 
\end{equation}
Note that for flat space, these expressions simplify as all terms
proportional to negative powers of $l$ vanish.

The results for AdS and flat space can be related through a modified
Penrose limit by translating the metric results of
\cite{Barnich:2012aw} to first order form. First, one introduces a
dependence on a dimensionless parameter $\epsilon>0$ in the arbitrary
functions $\Xi_{\pm\pm}$ of the AdS results giving rise to $\epsilon$
dependent vielbeins and spin connections
$e^{(\epsilon)},\omega^{(\epsilon)}$. If, after the rescalings
$(u,\phi,r;e^{(\epsilon)},\omega^{(\epsilon)})\to (\epsilon u,\epsilon
r,\phi;\epsilon^{-1}e^{(\epsilon)},\omega^{(\epsilon)})$, the limit is
well-defined it can be shown to be a solution to the flat space
equations. This is the case if the $\epsilon$ dependence\footnote{When
  explicitly comparing to \cite{Barnich:2012aw} one has to take into
  account the different normalization of the charges and the
  associated constant shifts of the functions
  $\Xi_{\pm\pm},\cM,\Theta$.}  in $\Xi_{\pm\pm}^{(\epsilon)}$ is such
that
\begin{equation}
\Xi_{\pm\pm}^{(\epsilon)}(x;\epsilon)=\frac{1}{4}\Theta(\pm x)\pm
\frac{\epsilon}{2l}\Xi(\pm x)+O(\epsilon^2),\label{eq:117}
\end{equation}
so that, when taking \eqref{eq:10} into account, 
\begin{equation}
  \label{eq:limit5}
\lim_{\epsilon\to 0}\mathcal{M}^{(\epsilon)}(\epsilon
 u,\phi)=\Theta(\phi),\quad 
\lim_{\epsilon\to 0}\epsilon^{-1} \mathcal{N}^{(\epsilon)}
(\epsilon
u,\phi)=\Xi(\phi)+\frac{u}{2}\partial_\phi\Theta(\phi).
\end{equation}

In the ${\rm AdS}$ case, the chiral
Chern-Simons connections are given on-shell by
\begin{equation}\label{connectionAdsnull}
A^{\pm} = \begin{pmatrix} \frac{r}{2l}dx^{\pm}&
\mp \frac{1}{\sqrt 2}\left (\frac{dr}{l} + \big(\frac{r^2}{2l^2}-2\Xi_{\pm \pm}
  \big)dx^\pm\right)\\ 
\pm \frac{1}{\sqrt 2}dx^\pm  & -\frac{r}{2l}dx^{\pm}\end{pmatrix} .
\end{equation}
in matrix form. 
They satisfy
\begin{equation}
\label{cbAds}
A^{\pm \alpha}_{\mp}=0,\quad A^{\pm 1}_r=0=A^{\pm 2}_r,\quad
\delta A^{\pm 0}_r=0=\d_\mu A^{\pm 0}_r. 
\end{equation}
Let us briefly compare to the formulation in the more standard
Fefferman-Graham gauge. In this case, the general solution is 
\begin{equation}\label{FG}
  ds^2=\frac{l^2}{r^2}dr^2-(r dx^+-\frac{l^2}{r}\Xi_{--}dx^-)
( r dx^--\frac{l^2}{r}\Xi_{++}dx^+),
\end{equation} 
where now $x^\pm=\frac{t}{l}\pm\phi$ with a standard time-like coordinate $t$
and a different radial coordinate $r$. Associated dreibeins $e^a$
and spin connections $\omega^a$ are
\begin{equation}
  \label{eq:24}
\begin{split}
  e^0=-\frac{r}{\sqrt 2} dx^-+\frac{l^2}{\sqrt 2 r}\Xi_{++}dx^+,
  e^1=\frac{r}{\sqrt 2} dx^+-\frac{l^2}{\sqrt 2r}\Xi_{--}dx^-,
  e^2=\frac{l}{r} dr, \\
  \omega^0=\frac{r}{\sqrt 2 l}dx^-+\frac{l}{\sqrt 2r}\Xi_{++}dx^+, 
\omega^1=\frac{r}{\sqrt 2
    l}dx^++\frac{l}{\sqrt 2r}\Xi_{--} dx^-,  \omega^2= 0.
\end{split}
\end{equation}
In this gauge as well, conditions \eqref{eq:18} hold.  The corresponding
chiral Chern Simons connections are 
\begin{equation}\label{connectionAdsnull2}
A^{+} = \begin{pmatrix} \frac{d r}{2
    r}&\frac{l}{r}\Xi_{++}dx^{+}\\ \frac{r}{l}dx^+  &
  -\frac{dr}{2r}\end{pmatrix},\  A^{-} = \begin{pmatrix}
  -\frac{d r}{2r}&\frac{r}{l}dx^{-}\\
  \frac{l}{r}\Xi_{--}dx^-  & \frac{dr}{2r} \end{pmatrix} . 
\end{equation}
In particular, these connections satisfy (i) $A^+_-=0=A^-_+$ and (ii)
$\d_\pm A^{+1}_+=0=\d_\pm A^{-0}_-$, $A^{\pm 2}_\pm=0$ for all values
of $r$, and thus in particular asymptotically, respectively to leading
order, which are the conditions used at spatial infinity in
\cite{Coussaert:1995zp}.

When comparing with \eqref{connectionAdsnull}, we see that (i) is
valid in both gauges, while (ii) is changed to $\d_+ A^{+
  0}_+=2\Xi'_{++}$, $\d_-A^{- 0}_-=-2\Xi'_{--}$, $\d_\pm
A^{+1}_+=0=\d_\pm A^{-1}_-$ and $A^{\pm 2}_\pm=0$ in the BMS gauge.

\section{On-shell group elements}
\label{sec:group-elements}

It is instructive at this stage to exhibit the group elements that
yield the flat connections discussed in the previous section. The
general solution to $dA+A^2=0$ is locally given by $A=G^{-1}dG$.

In the ${\rm AdS}$ case, we deduce from $F^\pm=0$ and $\d_\pm
A^{\pm}_r=0=\d_\mp A^\pm_r$ that $A^\pm =G_\pm^{-1} dG_\pm $ where
$G_\pm$ factorizes as $G_\pm=g_\pm (u,\phi)h_\pm(r)$.

In Fefferman-Graham gauge, the explicit form of the chiral
connections then leads to 
\begin{equation}
G_\pm=g_\pm e^{\pm \half \ln{\frac{r}{l}} H}\label{eq:17},
\end{equation}
with 
\begin{equation}
\begin{split}
  g_+^{-1}\d_+g_+ =\Xi_{++}  E_++ E_-,\quad \d_-
  g_+=0,\\ g^{-1}_-\d_- g_-=E_++ \Xi_{--} E_-,\quad  \d_+
  g_-=0\label{eq:22}.
\end{split}
\end{equation}
When using the Gauss parametrization
\begin{equation}
\label{Gauss}
g_\pm=e^{\sigma_\pm E_{\mp}}e^{\frac{1}{2}\mp \varphi_\pm H}e^{\tau_\pm E_\pm},
\end{equation}
this implies $\d_\mp\sigma_\pm=\d_\mp\tau_\pm=\d_\mp\varphi_\pm=0$
and
\begin{equation}
  \label{eq:27}
\d_\pm\sigma_\pm=e^{\varphi_\pm},\quad \tau_\pm\d_\pm\sigma_\pm
=-\half \d_\pm e^{\varphi_\pm},\quad 
\d_\pm\tau_\pm -\half \tau_\pm\d_\pm\varphi_\pm =\Xi_{\pm\pm}, 
\end{equation}
or, equivalently, in Riccatti form
\begin{equation}
  \label{eq:32}
  \d_\pm\tau_\pm +\tau^2_\pm=\Xi_{\pm\pm},
\quad \d_\pm\varphi_\pm=-2 \tau_\pm,\quad 
\d_\pm\sigma_\pm=e^{\varphi_\pm}. 
\end{equation}
When substituting the second in the first equation, one recognizes the
characteristic expression for the energy-momentum tensor of a
Liouville field, 
\begin{equation}
  \label{eq:120}
\Xi_{\pm\pm} =\frac{1}{4}(\d_\pm
\varphi_\pm)^2-\half \d_\pm^2\varphi_\pm 
= \d_\pm \tau_\pm+\tau_\pm^2= -\half \{\sigma_\pm;x^\pm\},
\end{equation}
where \(\{F;x\}=\frac{F'''}{F'}-\frac{3}{2}\frac{(F'')^2}{(F')^2}=(\ln
F')''-\half ((\ln F')')^2\) denotes the Schwarzian derivative. 
More precisely, consider a Liouville field $\varphi_L$ with action
\begin{equation}
  S=\int dud\phi ( \pi \dot
  \varphi_L-\half \pi^2 -\frac{1}{2 l^2}{\varphi_L'}^2
  -\frac{\mu}{2\gamma^2} e^{\gamma\varphi_L} )\label{eq:111}.
\end{equation}
It gives rise to the same relation (see e.g.~equations (4.3), (4.5),
(4.11) of \cite{Barnich:2012rz} for details) provided that the first
of \eqref{eq:27} holds, that \footnote{Note that
  $\sigma_+,\sigma_-$ are denoted by $A,B$ in section 4.1 of
  \cite{Barnich:2012rz}.} 
\begin{equation}
e^{\gamma\varphi_L}=\frac{16}{l^2\mu}
\frac{\d_+\sigma_+\d_-\sigma_-}{(\sigma_+-\sigma_-)^2}\label{eq:74}.
\end{equation}
and  
\begin{equation}
\gamma^2 l^2=32\pi G.\label{eq:112}
\end{equation}
For later use, let us point out that, off-shell, equation
\eqref{eq:74} together with an associated change of variables for the
momenta,
\begin{equation}
\begin{split}
 \label{eq:29}
  & \gamma
  \varphi_L=\varphi_++\varphi_--2\ln{(\sigma_+-\sigma_-)}+\ln{\frac{16}{l^2\mu}},\\
   & \pi=\frac{1}{\gamma
      l}\big(\varphi_{+}^{\prime}-\varphi_{-}^{\prime}-2\frac{\sigma_{+}^{\prime}+
\sigma^{\prime}_-}{\sigma_+-\sigma_-}\big),
\end{split}
\end{equation}
where $\sigma_\pm'=\pm e^{\varphi_\pm}$, is the change of variables that
allows to write the Liouville action \eqref{eq:111} in terms of
decoupled chiral bosons,
\begin{equation}
\label{limit:10c}
S=\frac{1}{\gamma^2 l}\int dud\phi
\Big[\dot{\varphi}_+\varphi'_+ -
\dot{\varphi}_-\varphi'_--\frac{1}{l}(\varphi_+^{\prime})^2-\frac{1}{l}(\varphi_-^\prime)^2\Big].
\end{equation}

In ${\rm BMS}$ gauge, one finds 
\begin{equation}
G_\pm=f_\pm(u,\phi)e^{\mp\frac{r}{l} j_0}.\label{eq:1}
\end{equation}
with 
\begin{equation}
  \label{eq:25}
  f^{-1}_\pm\d_\pm f_\pm=\pm \sqrt 2\Xi_{\pm\pm} E_+\pm
  \frac{1}{\sqrt 2} E_-,\quad \d_\mp f_\pm =0.
\end{equation}
In this case, the parametrization 
\begin{equation}
  \label{eq:28}
  f_{\pm}=e^{\pm\frac{\sigma_\pm}{\sqrt 2}
    E_{-}}e^{-\frac{1}{2}\varphi_\pm H}e^{\pm\sqrt 2 \tau_\pm E_+}, 
\end{equation}
leads to the same equations \eqref{eq:27} or \eqref{eq:32} as in
Fefferman-Graham gauge.

Finally, in the flat case, we start by solving $d\omega+\half
\omega^2=0$. Since all $r$ and $u$ dependence drops out of
\eqref{eq:12} and 
\begin{equation}
\omega=\frac{1}{2\sqrt 2}\Theta d\phi
E_++\frac{1}{\sqrt 2}d\phi E_-,\label{eq:98}
\end{equation}
we have $\omega=\Lambda^{-1} d\Lambda$ where
$\Lambda=\Lambda(\phi)$. The parametrization 
\begin{equation}
  \label{eq:30}
 \Lambda=e^{\frac{\sigma}{\sqrt 2}
    E_{-}}e^{-\frac{1}{2}\varphi H}e^{\sqrt 2 \tau E_+}
\end{equation}
then leads to 
\begin{equation}
  \label{eq:31}
  \sigma'=e^{\varphi},\quad \tau\sigma'
=-\half\varphi' e^{\varphi},\quad 
\tau' -\half \tau\varphi' =\frac{1}{4}\Theta, 
\end{equation}
or, again, in Riccatti form
\begin{equation}
  \label{eq:32a}
  \tau'+\tau^2=\frac{1}{4}\Theta,
\quad \varphi'=-2\tau,\quad 
\sigma'=e^{\varphi}. 
\end{equation}
In this case, we have 
\begin{equation}
  \label{eq:120a}
\frac{1}{4}\Theta =\frac{1}{4}\varphi'^2-\half \varphi''
=\tau'+\tau^2= -\half \{\sigma;\phi\}.
\end{equation}

The equation $de+\omega e+ e\omega=0$ is locally solved by
$e=\Lambda^{-1} da \Lambda$. 
With 
\begin{equation}
e=\big[\frac{1}{2} \Theta du -dr +(\Xi +\frac{u}{2} \d_\phi
\Theta)d\phi\big]\frac{1}{\sqrt 2} E_+ +\half r d\phi H + du
\frac{1}{\sqrt 2} E_-,\label{eq:37}
\end{equation}
the ansatz 
\begin{equation}
\label{limit:7}
\begin{split}
a=-\frac{r}{\sqrt{2}} \Lambda E_+\Lambda^{-1} +u \d_{\phi}\Lambda
\Lambda^{-1}+\bar a(\phi),\\ 
\bar a(\phi) =\frac{\eta}{\sqrt 2} E_{+}+\frac{\theta}{2}
H+\frac{\zeta}{\sqrt 2} E_-,
\end{split}
\end{equation}
then leads to the system
\begin{equation}
  \label{eq:119}
  \eta'={e^{-\varphi}}\Xi,\quad  \theta'= -{\sigma  e^{-\varphi}}\Xi,
  \quad \zeta'=- \frac{\sigma^2}{2}e^{-\varphi}\Xi,
\end{equation}
which can be trivially integrated. 

Let us compare with a BMS Liouville field \cite{Barnich:2012rz}
with action
\begin{equation}
  \label{eq:113}
  S=\int du d\phi ( \Pi\dot\Phi -\half {\Phi'}^2 -\frac{\nu}{2\beta^2}
  e^{\beta\Phi}). 
\end{equation}
Taking into account the expression for the energy density of this
field leads to the relation
\begin{equation}
e^{\beta\Phi}=\frac{4}{\nu}(\frac{\sigma'}{\sigma})^2,\label{eq:97}
\end{equation}
where the first of \eqref{eq:31} holds, by using\footnote{The function
  $\sigma$ is denoted by $B$ in section 4.3 of \cite{Barnich:2012rz}.}
the first of equations (4.29) and (4.30) of
\cite{Barnich:2012rz}. From (4.22) of \cite{Barnich:2012rz} and
\eqref{eq:110}, we then get
\begin{equation}
  \label{eq:114}
  \beta^2=32\pi G. 
\end{equation}
Off-shell, the change of variables from $\Phi,\Pi$ to
$\varphi,\xi$ given by
\begin{equation}
  \label{eq:38}
\begin{split}
  &  \beta\Phi=2\varphi-2\ln{\sigma}+\ln{\frac{4}{\nu}},\\
  & \beta \Pi=\xi'-(\ln{\sigma})'\xi,
\end{split}
\end{equation}
where $\sigma'=e^\varphi$ maps the BMS Liouville action \eqref{eq:113}
to
\begin{equation}
  \label{eq:42}
  S=\frac{2}{\beta^2}\int du d\phi\Big[\xi' \dot \varphi
  -\varphi'^2\Big]. 
\end{equation}

We are now in a position to discuss the flat limit. On the level of the
Liouville action \eqref{eq:111},  let
\begin{equation}
\begin{split}
& \varphi_L=\epsilon^{-\half} l \Phi,\quad 
\pi=\epsilon^{\half} l^{-1} \Pi ,\\ 
& \gamma=\epsilon^{\half}
l^{-1}\beta ,\quad \mu= l^{-2} \nu.\label{eq:53}
\end{split}
\end{equation}
After rescaling $u\to \epsilon u$, the limit $\epsilon\to 0$ of
\eqref{eq:111} then gives rise to the BMS Liouville action
\eqref{eq:113}.

On the level of solutions, consider the group element
$f^{(\epsilon)}_\pm(x;\epsilon)$ determined through \eqref{eq:25} where
$\Xi_{\pm\pm}$ is replaced by $\Xi^{(\epsilon)}_{\pm\pm}$ with an
expansion as in \eqref{eq:117}. A parametrization like in
\eqref{eq:28} with $\epsilon$ dependent fields
$\sigma_\pm^{(\epsilon)},\varphi_\pm^{(\epsilon)},\tau_\pm^{(\epsilon)}$
now gives rise to equations \eqref{eq:32}, \eqref{eq:120} in terms of
$\epsilon$ dependent fields. Compatibility with \eqref{eq:32a},
\eqref{eq:120a}, \eqref{eq:119} then leads to
\begin{equation}
\label{limit:1}
\begin{split}
  & \tau_{\pm}(x;\epsilon)=\pm\tau(\pm x)\pm \epsilon \tau^{(1)} (\pm
  x) +O(\epsilon^2),\  l\tau^{(1)}=\frac{1}{2}
  e^{\varphi}\eta, \\
  & \varphi_{\pm}(x;\epsilon)=\varphi(\pm x) 
  +\epsilon\varphi^{(1)}(\pm x)+ O(\epsilon^2),\ l\varphi^{(1)}=-\theta-\eta\sigma,\\
  & \sigma_{\pm}(x;\epsilon)=\pm\sigma(\pm x)\pm \epsilon
  \sigma^{(1)}(\pm x)+O(\epsilon^2),\ 
l\sigma^{(1)}=\zeta-\theta\sigma-\half \eta \sigma^2. 
\end{split}
\end{equation} 

For the chiral groups elements we find
\begin{eqnarray}
\label{limit:2}
f_\pm(x;\epsilon) = \big[\Lambda(\mathbb 1+\epsilon \frac{b_{\pm}}{l})\big](\pm x)
+O(\epsilon^2),
\end{eqnarray}
with $\Lambda$ parametrized as in \eqref{eq:30} and
\begin{eqnarray}
\label{limit:3}
\frac{b_{\pm}}{l}=\frac{\sigma^{(1)}}{\sqrt{2}}e^{-\varphi}(E_--\sqrt 2
\tau H-2 \tau^2 E_+)
+\sqrt{2}\left(\tau^{(1)} -\tau \varphi^{(1)}\right) E_+
-\half\varphi^{(1)}  H.
\end{eqnarray}
This gives
\begin{equation}
\label{limit:5}
f^{-1}_\pm(x;\epsilon)\d f_\pm(x;\epsilon)= \pm \Big[\Lambda^{-1}\d\Lambda
+ \frac{\epsilon}{l}\big(\d b_\pm
+[\Lambda^{-1}\d \Lambda,b_\pm ]\big)\Big] (\pm x)
+O(\epsilon^2),
\end{equation}
According to the discussion on the flat limit in Section
\bref{sec:bms-gauge-boundary}, one defines
\begin{equation}
  G^\epsilon_\pm=f_\pm(\frac{\epsilon u}{l}\pm\phi;\epsilon) e^{\mp
    \frac{\epsilon r}{l\sqrt 2} E_+},\label{eq:118}
\end{equation}
and computes
\begin{equation}
\label{limit:6}
 \omega=
\lim_{\epsilon \to 0} \frac{1}{2}\big({G^\epsilon_{+}}^{-1} 
d G^\epsilon_{+}+{G^\epsilon_{-}}^{-1} d
  G^\epsilon_{-}\big),\,
e=\lim_{\epsilon \to 0}\frac{l}{2\epsilon}\big({G^\epsilon_{+}}^{-1}
  d G^\epsilon_{+}-{G^\epsilon_{-}}^{-1}d G^\epsilon_{-}\big).
\end{equation}
This reproduces the flat space results discussed previously, that is 
 $\omega= \Lambda^{-1} d \Lambda$ with $\Lambda=\Lambda(\phi)$
and $e=\Lambda^{-1} d a \Lambda$ with $a$ given by 
\begin{equation}
\label{limit:7b}
a=-\frac{r}{\sqrt{2}} \Lambda E_+\Lambda^{-1} +u \d_{\phi}\Lambda 
\Lambda^{-1}+\half \Lambda (b_++b_-)\Lambda^{-1},
\end{equation}
by taking into account that  $\half (b_++b_-)=\Lambda^{-1}\bar
a\Lambda$. 

\section{Improved action principle}
\label{sec:reduct-wess-zumino}

Neglecting again boundary terms and using $A=du A_u+\tilde A$,
$d=du\partial_u +\tilde d$, the Chern-Simons action can directly be
written in Hamiltonian form, 
\begin{equation}
  \label{eq:16}
\begin{split}
  S[A]
& =-\frac{k}{2\pi} \int du dr d \phi \left( \epsilon^{ij}
    \omega_{a i}\ \d_u
    e^a_j-\mathcal H \right),\\
\gamma^a_e & =\frac{1}{2}\epsilon^{ij}[\partial_{i}e^{a}_j
-\partial_{j}e^{a}_i+\epsilon^{abc}(\omega_{ib}e_{jc}-
\omega_{jb}e_{ic})],\\
\gamma^a_\omega & =\frac{1}{2}\epsilon^{ij}[\partial_{i}\omega^{a}_j-
\partial_{j}\omega^{a}_i+\epsilon^{abc}(\omega_{ib}
\omega_{jc}+\frac{1}{l^2}e_{ib}e_{jc})], 
\end{split}
\end{equation}
where the Hamiltonian density is a combination of constraints,
$\mathcal H= e_{u a} \gamma^a_\omega +\omega_{u a}
\gamma^a_e$ and where $x^i= r,\phi$ with $\epsilon^{ij}$
determined by $\epsilon^{12}=-1$.

At this stage, we have to discuss boundary terms. On shell, 
\begin{equation}
  \label{eq:19}
  \delta \left( \epsilon^{ij}  \omega_{a i}\ \d_u e^a_j
 - \mathcal H\right)=-\d_i\left ( \epsilon^{ij} (e_{ua}\delta
  \omega^a_j+ \omega_{ua}\delta e^a_j)\right) +\d_u
\left( \epsilon^{ij} \omega_{a i} \delta e^a_j\right). 
\end{equation}
It follows that 
\begin{multline}
  \label{eq:20}
  \delta\ \int dudrd\phi \left( \epsilon^{ij}
    \omega_{a i}\ \d_u
    e^a_j-\mathcal H\right)= -\int du d\phi \Big[
    e_{ua}\delta \omega^a_\phi+\omega_{ua}\delta
      e^a_\phi\Big]^{r=\infty}_{r=\bar r} -\\- \int du dr
    \Big [ e_{ua}\delta \omega^a_r+\omega_{ua}\delta
        e^a_r\Big]^{\phi=2\pi}_{\phi=0}
-\int drd\phi\Big[\omega_{a r}\delta
  e^a_\phi-\omega_{a\phi} \delta
  e^a_r\Big]^{u=u_f}_{u=u_i}. 
\end{multline}
Assuming the fields to be single-valued on the circle, neglecting the
inner boundary at $r=\bar r$ and taking into account conditions
\eqref{eq:18}, solutions \eqref{eq:12} respectively \eqref{eq:24}
provide a true extremum of the variational principle defined by
\begin{multline}
  \label{eq:21}
  I[e,\omega]=-\frac{k}{2\pi} \int dudrd\phi \left( \omega_{a\phi}\dot
    e^a_r-
\omega_{a r}\dot
    e^a_\phi-\mathcal
    H\right)-\\-\frac{k}{4\pi}\int du d\phi \Big[\omega_{a\phi}
  \omega^a_\phi+\frac{1}{l^2}
  e_{a\phi}e^a_\phi\Big]^{r=\infty}.
\end{multline}

\section{Reduction to Wess-Zumino-Witten theory}
\label{sec:reduct-wess-zumino-1}

In all cases, ${\rm BMS}$ gauge in ${\rm AdS}$ and flat space and in
Fefferman-Graham gauge in ${\rm AdS}$ space, the on-shell vielbeins
and spin connections satisfy in particular
\begin{equation}
\d_\phi e_r=0=\d_\phi \omega_r=0\label{eq:40},
\end{equation}
which are a subset of conditions \eqref{eq:18}. In an off-shell
formulation, these conditions can be taken as (partial) gauge fixing
conditions. The reduced system is then simply obtained by solving the
constraints with those gauge fixing conditions in the action.

Because of the form of the constraints, the analysis has to be done
separately depending on whether there is a cosmological constant or
not. In the next subsection we briefly review the results in the AdS
case along the lines of \cite{Coussaert:1995zp} in order to better
appreciate what happens in the flat case. Standard technical material
can be found in the appendix.

\subsection{AdS case}

In the ${\rm AdS}$ case, one uses the chiral decomposition in terms of which
the constraints split as 
\begin{equation}
  \label{eq:23}
  \tilde d \tilde A^{\pm}+(\tilde A^{\pm})^2=0,
\end{equation}
and the improved action is
\begin{equation}
  \label{eq:45}
  I[e,\omega]= I^c[A^+]-I^c[A^-]-\frac{k}{4\pi} \int dud\phi\ {\rm
    Tr}\Big[(A^+_\phi)^2+(A^-_\phi)^2\Big]^{r=\infty} ,
\end{equation}
where
\begin{equation}
  \label{eq:47}
  I^c[A]=  -\frac{k l}{4\pi} \int dudrd\phi\ {\rm Tr} \Big(
 A_\phi\dot A_r -A_r \dot A_\phi\Big) - \frac{k l}{4\pi}\int du\ {\rm Tr}\Big( A_u
  \big(\tilde d \tilde A+\tilde A^2\big)\Big). 
\end{equation}

The general solution to the constraints is locally given by $\tilde
A^\pm= G^{-1}_\pm \tilde dG_\pm$.  Taking into account in addition the
gauge fixing conditions, which can be rewritten as $\d_\phi
A^\pm_r=0$, the general solution factorizes,
\begin{equation}
G_\pm=g_\pm(u,\phi)h_\pm(r,u),\label{eq:42a}
\end{equation}
in terms of group elements
$g_\pm,h_\pm$. At finite $r=\bar r$ fixed, we can assume without loss
of generality that $\dot h_\pm(\bar r,u)=0$ by absorbing $h_\pm(\bar
r,u)$ into $g_\pm(u,\phi)$. We will assume that this condition also
holds for $\bar r=\infty$, 
\begin{equation}
  \label{eq:121}
  \dot h_\pm(\infty,u)=0,
\end{equation}
as it indeed does for the gravitational solutions of interest.

When inserting such a solution of the constraints into the improved
action, one gets the sum of two decoupled chiral Wess-Zumino-Witten
models, $I[e,\omega]=I_+[g_+]+I_-[g_-]$,
\begin{equation}
  \label{eq:15}
  I_\pm[g_\pm]=\pm\frac{k}{2\pi} \int du d\phi {\rm Tr}
  \Big[(g^{-1}_\pm\d_+ g_\pm-g^{-1}_\pm\d_- g_\pm)g^{-1}_\pm\d_\mp g_\pm\Big] \pm \frac{kl}{2\pi}
    \Gamma[G_\pm],
\end{equation}
where 
\begin{equation}
  \label{eq:46}
 \Gamma[G]=\frac{1}{3!}\int {\rm Tr} (G^{-1}dG)^3.
\end{equation}

One can then follow \cite{Coussaert:1995zp} and define
\begin{equation}
G=G_+^{-1}G_-,\quad g=g_+^{-1}g_-,\quad \pi= -g_-^{-1}
g_+'g_+^{-1}g_--g_-^{-1} g_-'\label{eq:36}, 
\end{equation}
in terms of which
\begin{multline}
  \label{eq:26bis}
  I[e,\omega]=\frac{ k l}{2\pi}\int du d\phi\ {\rm
    Tr}\Big[\half \pi g^{-1} \dot g -\frac{1}{4l}\big(\pi^2+(g^{-1}
  g')^2\big)\Big]
- \frac{k l}{2\pi} \Gamma[G].
\end{multline} 
After elimination of the momenta, one gets the (non-chiral) WZW
action at the boundary $r=\infty$, 
\begin{equation}
  \label{eq:34}
   I[g]  =-\frac{k l^2}{8\pi}\int du d\phi\ {\rm
    Tr}\Big[\eta^{\mu\nu}  g^{-1} \d_\mu g g^{-1} \d_\nu g\Big]
- \frac{k l}{2\pi}\Gamma[G] .
\end{equation}

The equations of motions of the non-chiral theory are $\d_+(g^{-1}\d_-
g)=0$, with general solution $g=k_+(x^+)k_-(x^-)$. It then follows
directly from the Polyakov-Wiegmann identities that the WZW action is
invariant under $g\to \Theta_+(x^+) g \Theta^{-1}_-(x^-)$. The
conserved Noether currents for the associated infinitesimal
transformations $\delta_{\theta_\pm}g=\theta_+ g-g\theta_-$ are
\begin{equation}
\label{eq:35}
J^+_{\theta_\pm} = -\frac{k}{\pi}{\rm Tr}\Big[\theta_-
g^{-1}\d_- g\Big], \quad
J^-_{\theta_\pm}  = \frac{k}{\pi} {\rm Tr}\Big[\theta_+
\d_+ g g^{-1}\Big],
\end{equation}
with time components 
\begin{equation}
J^0_{\theta^\pm}=2{\rm Tr}[\theta_\pm
I_\pm],\quad  I_+=\frac{kl}{4\pi}\d_+ gg^{-1},\quad 
  I_-= -\frac{kl}{4\pi}g^{-1}\d_- g.\label{eq:101}
\end{equation}
As briefly recalled in the appendix, in the Hamiltonian formulation
their Poisson bracket algebra consists of two commuting copies of an
$\mathfrak{sl}(2,\mathbb R)$ current algebra,
\begin{equation}
  \label{eq:74b}
\begin{split}
&  \{I^\pm_a(\phi),I^\pm_b(\phi^\prime)\}=
  {\epsilon_{ab}}^c I^\pm_c(\phi)
\delta(\phi-\phi^\prime)\pm
\frac{k l}{4\pi }\eta_{ab}\d_\phi\delta(\phi-\phi^\prime), \\
 & \{I^+_a(\phi),I^-_b(\phi^\prime)\}=0.
\end{split}
\end{equation}

Alternatively, in order to better compare with the flat case, one can
concentrate on the two chiral copies. The equations of motion are
$\d_\mp (g_\pm^{-1}g_\pm')=0$ which implies $g_\pm =h_\pm (u)
k_\pm(x^\pm)$. The analog of the Polyakov-Wiegmann identities for the
chiral case imply invariance of the chiral theories $I_\pm[g_\pm]$
under $g_\pm \to g_\pm \Theta^{- 1}_\pm (x^\pm)$.  The conserved
Noether currents for the associated infinitesimal transformations
$\delta_{\theta_\pm }g_\pm=- g_\pm\theta_\pm$ are given by
\begin{equation}
  \label{eq:43}
  \begin{split}
  &  J^+_{\theta_+}=0,\quad J^-_{\theta_+}=-\frac{k}{\pi}{\rm Tr}
    \Big[\theta_+ g^{-1}_+g'_+\Big],\\
& J^+_{\theta_-}=\frac{k}{\pi}{\rm Tr}\Big[\theta_- g^{-1}_-
g'_-\Big],\quad J^-_{\theta^-}=0.
  \end{split}
\end{equation}
The time-components of these currents can be written as 
\begin{equation}
J^0_{\theta_\pm}=2{\rm Tr}[\theta_\pm I^\pm],\quad 
I^\pm=\mp \frac{kl}{4\pi} g_\pm^{-1} g'_\pm.
\label{eq:43a}
\end{equation}
They agree on-shell with the current components \eqref{eq:101} of the
non chiral WZW theory, which justifies the same notation.  In this
case, it is their Dirac bracket algebra that forms the $\mathfrak{sl}(2,\mathbb
R)$ current algebra given in \eqref{eq:74b}. This is shown in the
appendix along the lines of \cite{Salomonson:1988mk}. The chiral
models are more complicated than the non-chiral theory in the sense
that their Hamiltonian formulation involves constraints, the zero
modes of which are first class and generate the arbitrary functions of
time in the general solution to the equations of motion, while all
other modes are second class.

Let us briefly recall how classical conformal invariance is expressed
on the level of the Hamiltonian formulation of the chiral models. On
the constraint surface, the Hamiltonian and momentum densities can be
written as
\begin{equation}
  \label{eq:56}
\begin{split}
&  \cH^\pm =\half \big[ \frac{1}{l^2} \frac{8\pi}{k}
(\pi^B_\pm)_a  (\pi^B_\pm)^a+\frac{k}{8\pi}(g^{-1}_\pm
g'_\pm)_a (g^{-1}_\pm g'_\pm)^a \big]\approx
\frac{2\pi}{kl^2}I^\pm_a I^a_\pm,\\
& \cP^\pm= -(\pi^B_\pm)^a (g^{-1}_\pm g'_\pm)_a\approx
\pm\frac{2\pi}{kl} I^\pm_a I^a_\pm.
\end{split}
\end{equation}
By using the Dirac bracket version of
\eqref{eq:74b}, one gets by direct computation, 
\begin{equation}
  \label{eq:63}
  \begin{split}
    \{\cH^\pm(\phi),\cH^\pm(\phi')\}^*=\frac{1}{l^2}(\cP^\pm(\phi)+\cP^\pm(\phi'))\d_\phi\delta(\phi-\phi'),\\
\{\cH^\pm(\phi),\cP^\pm(\phi')\}^*=(\cH^\pm(\phi)+\cH^\pm(\phi'))\d_\phi\delta(\phi-\phi'),\\
\{\cP^\pm(\phi),\cP^\pm(\phi')\}^*=(\cP^\pm(\phi)+\cP^\pm(\phi'))\d_\phi\delta(\phi-\phi').
  \end{split}
\end{equation}
For the energy-momentum tensor components
$T^\pm_{uu}=\cH^\pm=\frac{1}{l^2}T^\pm_{\phi\phi}$,
$T^\pm_{u\phi}=\cP^\pm=T^\pm_{\phi u}$, the components in light-cone coordinates
\begin{equation}
\begin{split}
& T_{++}^\pm=\frac{l^2}{2}(\cH^\pm+\frac{1}{l}\cP^\pm)\approx
\delta_{+}^\pm\frac{2\pi}{k}I^+_a I^a_+,\\ 
& T_{--}^\pm=\frac{l^2}{2}(\cH^\pm-\frac{1}{l}\cP^\pm)\approx
\delta_{-}^\pm\frac{2\pi}{k}I^-_a I^a_-,\\
& T_{+-}^\pm=0=T_{-+}^\pm,\label{eq:57}
\end{split}
\end{equation}
satisfy
\begin{equation}
  \label{eq:59}
  \{T_{\pm\pm}^\pm(\phi),T_{\pm\pm}^\pm(\phi')\}^*=\pm l (T_{\pm\pm}^\pm(\phi)
+T_{\pm\pm}^\pm(\phi'))\delta'(\phi-\phi'),
\end{equation}
and all other brackets vanishing.

\subsection{Flat case}
\label{sec:flat-case}

In this case, taking into account that $\d_\phi\omega_r=0$, the
general solution to $\gamma_\omega=0$ is given by $\tilde \omega=
\Lambda^{-1}\tilde d\Lambda$ where $\Lambda=\lambda(u,\phi)\mu(r,u)$,
and $\lambda,\mu$ are group elements.  The general solution to the
remaining constraint $\gamma_e=0$, which is equivalent to $d\tilde
e+\tilde\omega\tilde e+\tilde e\tilde\omega=0$, together with the
gauge fixing condition $\d_\phi e_r=0$, is given by
\begin{equation}
  \label{eq:49}
  \tilde e=\Lambda^{-1} \tilde d a\Lambda, \quad a=\alpha+\lambda \beta
  \lambda^{-1},
\end{equation}
where $\alpha=\alpha(u,\phi)$, $\beta=\beta(u,r)$, which gives
explicitly 
\begin{equation}
  \label{eq:50}
\begin{split}
  & \omega_r = \mu^{-1} \d_r \mu,\quad
  \omega_\phi= \mu^{-1}\lambda^{-1}\lambda'\mu,\\
& e_r=\mu^{-1} \d_r\beta \mu,\quad e_\phi= 
\mu^{-1}\big(\lambda^{-1} 
\alpha'\lambda+[\lambda^{-1}\lambda',\beta]\big)\mu. 
\end{split}
\end{equation}

Inserting this solution into the improved action \eqref{eq:21} with
$l\to \infty$ gives  
\begin{equation}
  \label{eq:51}
  I[e,\omega]= \frac{k}{\pi} \int dud\phi\, {\rm Tr}\, \Big[
  \Lambda'\Lambda^{-1}\dot a-\half (\Lambda^{-1}
  \Lambda')^2 \Big]+
  \frac{k}{\pi}\Gamma[\Lambda,a], 
\end{equation}
with 
\begin{equation}
  \label{eq:52}
  \Gamma[\Lambda,a]  =\int {\rm Tr}\, \big( d\Lambda \Lambda^{-1}
  d\Lambda \Lambda^{-1} d a\big).
\end{equation}
As for the non semi-simple Lie algebra considered in
\cite{Nappi:1993ie}, this Wess-Zumino term for the
Poincar\'e algebra $\mathfrak{iso}(2,1)$ is exact, 
\begin{equation}
  \label{eq:39}
   {\rm Tr}\, \big( d\Lambda \Lambda^{-1}
  d\Lambda \Lambda^{-1} d a\big)=d \Big[{\rm Tr}
  \big(d\Lambda \Lambda^{-1}d a\big)\Big],
\end{equation}
so that, when concentrating on the boundary at $r=\infty$, 
\begin{equation}
  \label{eq:54}
 \Gamma[\Lambda,a]=\int dud\phi\, {\rm Tr}\, \Big[\dot
  \Lambda\Lambda^{-1}a' -\Lambda'\Lambda^{-1} \dot
  a\Big].
\end{equation}
Furthermore, when using the decompositions of $\Lambda$ and $a$, 
\begin{multline}
{\rm Tr}\Big[\dot{\Lambda} \Lambda^{-1}a'\Big]
={\rm Tr}\Big[\dot{\lambda} \lambda^{-1}\alpha'
+\dot{\mu}
\mu^{-1}(\lambda^{-1}\alpha'
 \lambda+\lambda^{-1
}\lambda' \beta  -\beta \lambda^{-1} \lambda')\\
+ \lambda^{-1}\lambda' \dot{\beta}
-\d_{u}(\lambda^{-1}\lambda'\beta)+
\d_{\phi}(\lambda^{-1}\lambda' \dot{\beta}) \Big].
\end{multline}
For the gravitational solutions of interest we have again
\begin{equation}
\dot
\mu(\infty,u)=0 =\dot \beta(\infty,u),\label{eq:122}
\end{equation}
so that only the first term survives and
\begin{equation}
  \label{eq:54a}
\boxed{  I[\lambda,\alpha]=\frac{k}{\pi} \int dud\phi\ {\rm Tr} \Big[ \dot
  \lambda\lambda^{-1}\alpha'-
  \half (\lambda'\lambda^{-1} )^2  \Big] }.
\end{equation}
This action differs from the WZW action for flat gravity proposed in
\cite{Salomonson:1989fw} by the potential energy term, which
originates from the boundary term in \eqref{eq:21}.

The equations of motion are 
\begin{equation}
 (\dot\lambda\lambda^{-1})'=0,\quad 
D_u^{-\dot\lambda\lambda^{-1}}\alpha^\prime=(\lambda'\lambda^{-1})',\label{eq:44}
\end{equation}
These equations are equivalent to the conservation laws $\partial_\mu
\tilde J^\mu=0$,
$\partial_\mu P^\mu=0$ where 
\begin{equation}
  \label{eq:83}
\tilde  J^0=\lambda^{-1} \alpha'\lambda,\quad
\tilde  J^1=-\lambda^{-1}\lambda',\quad P^0=\lambda'\lambda^{-1},\quad P^1=0.
\end{equation}

The general solution of the first equation is
$\lambda=\mu(u)\nu(\phi)$. After defining $\alpha=\mu \gamma \mu^{-1}$
the second equation reads $\dot\gamma'=(\nu'\nu^{-1})'$ with general
solution $\gamma=\rho(\phi)+\delta(u) + u \nu'\nu^{-1}$. 

Solution space is invariant under
$\lambda\to\lambda\Theta^{-1}(\phi)$, $\alpha\to \alpha -
u\lambda\Theta^{-1}\Theta'\lambda^{-1}$ and also under $\lambda\to
\Xi(u)\lambda$, $\alpha \to \Xi\alpha\Xi^{-1}$.  The infinitesimal
version of the former, $\delta_\theta\lambda=-\lambda\theta$,
$\delta_\theta\alpha =-u \lambda\theta'\lambda^{-1}$, leave action
\eqref{eq:54a} invariant and the associated Noether currents are now
$\tilde J^0_\theta=-\frac{k}{\pi} {\rm Tr}[u
\theta'\lambda^{-1}\lambda'+\theta\lambda^{-1}\alpha'\lambda]$,
$\tilde J^1_\theta=\frac{k}{\pi} {\rm Tr}[\theta\lambda^{-1}\lambda']$.  For
Noether currents, the physically meaningful quantity is the
equivalence class $[J^\mu]$, where $J^\mu\sim J^\mu+t^\mu+\d_\nu
k^{[\mu\nu]}$ with $t^\mu\approx 0$. Choosing $k^{[\mu\nu]}=\frac{k}{\pi} u
\epsilon^{\mu\nu}{\rm Tr}(\theta \lambda^{-1}\lambda')$ and
$t^\mu=\frac{k}{\pi} \delta^\mu_1{\rm Tr}\big[\theta(\d_0
(\lambda^{-1}\lambda')\big]$, an equivalent representative for the
Noether current is
\begin{equation}
  \label{eq:96}
  J^0_\theta=2{\rm Tr}[\theta J],\quad
J=-\frac{k}{2\pi}\big[\lambda^{-1}\alpha'
\lambda-u(\lambda^{-1}\lambda')'\big],\quad
  J^1_\theta=0. 
\end{equation}
The action is furthermore invariant under $\lambda\to\lambda$ and
$\alpha\to \alpha+\lambda \Sigma(\phi)\lambda^{-1}$. Associated infinitesimal
transformations are $\delta_\sigma\lambda=0, \delta_\sigma
\alpha=\lambda \sigma \lambda^{-1}$ with Noether currents
\begin{equation}
  \label{eq:85}
  P^0_\sigma=2{\rm Tr}[\sigma P],\quad P=\frac{k}{2\pi} 
\lambda^{-1}\lambda',\quad
  P^1_\sigma=0.
\end{equation}
As shown in the appendix, in the Hamiltonian formulation, the Dirac
brackets of their time components satisfy the $\mathfrak{iso}(2,1)$
current algebra
\begin{equation}
  \label{eq:94}
  \begin{split}
& \{P_a(\phi),P_b(\phi')\}^*=0,\\
& \{J_a(\phi),P_b(\phi')\}^*={\epsilon_{ab}}^c
P_c(\phi)\delta(\phi-\phi')-\frac{k}{2\pi}
\eta_{ab}\d_\phi\delta(\phi-\phi'),\\
 &   \{J_a(\phi),J_b(\phi')\}^*={\epsilon_{ab}}^c
    J_c(\phi)\delta(\phi-\phi').
  \end{split}
\end{equation}

Let us now discuss BMS3 invariance of the model. Using the Hamiltonian
analysis that has been done in the appendix, it follows that, on the
constraint surface, the Hamiltonian and momentum densities are given
by
\begin{equation}
  \label{eq:61}
  \cH\approx \frac{\pi}{k}P^a P_a,\quad \cP\approx -\frac{2\pi}{k}
  J^a P_a. 
\end{equation}
When using the current algebra \eqref{eq:94}, we find
\begin{equation}
  \label{eq:62}
  \begin{split}
& \{\cH(\phi),\cH(\phi')\}^*=0,\\
& \{\cH(\phi),\cP(\phi')\}^*=(\cH(\phi)+\cH(\phi'))\d_\phi\delta(\phi-\phi'),\\
& \{\cP(\phi),\cP(\phi')\}^*=(\cP(\phi)+\cP(\phi'))\d_\phi\delta(\phi-\phi').
  \end{split}
\end{equation}
This is the form BMS3 invariance takes in the Hamiltonian
framework. Indeed, in terms of modes $P_m=\int^{2\pi}_0 d\phi\ e^{im\phi} \cH$,
$J_m=\int^{2\pi}_0 d\phi\ e^{im\phi} \cP$, one finds 
\begin{equation}
  \label{eq:64}
     i\{P_m,P_n\}=0,\quad  i\{J_m,P_n\}=(m-n) P_{m+n},\quad
     i\{J_m,J_n\}=(m-n) J_{m+n}. 
\end{equation}

When translating to the Lagrangian level, the transformations  
\begin{equation}
  \label{eq:65}
-\delta_\xi=\{\cdot,Q_\xi\},\quad  Q_\xi=\int^{2\pi}_0 d \phi ( \cH T
+\cP Y) , 
\end{equation}
where $T=T(\phi)$, $Y=Y(\phi)$, are expressed through
\begin{equation}
  \label{eq:66}
  -\delta_\xi \lambda=Y\lambda',\quad -\delta_\xi\alpha
  =f\lambda'\lambda^{-1} +Y\alpha',\quad f= T+u Y'. 
\end{equation}
It can then readily be checked that they leave action \eqref{eq:54a}
invariant, $\delta_\xi \cL=\d_\mu k^\mu_\xi$ and that the
energy-momentum tensor 
$-j^\mu_\xi\equiv
{T^\mu}_\nu\xi^\nu=k^\mu_\xi+\ddl{\cL}{\d_\mu\lambda}\delta_\xi
\lambda+  \ddl{\cL}{\d_\mu\alpha}\delta_\xi
\alpha$ reads
\begin{equation}
  \label{eq:80}
  {T^u}_\nu \xi^\nu=-\frac{k}{\pi}{\rm
    Tr}\big[\half(\lambda'\lambda^{-1})^2
  f+\lambda'\lambda^{-1}\alpha' Y\big], \quad {T^\phi}_\nu \xi^\nu=\frac{k}{2\pi}{\rm
    Tr}\big[(\lambda'\lambda^{-1})^2Y\big]. 
\end{equation}
Agreement with the Hamiltonian analysis follows by using $\tilde
  {T^\mu}_\nu \xi^\nu ={T^\mu}_\nu\xi^\nu+\d_\rho k^{[\rho\mu]}_\xi$, with
$k^{[u\phi]}_\xi=-\frac{k}{2\pi}{\rm Tr}\big[u(\lambda'\lambda^{-1})^2Y\big]$,
so that 
\begin{equation}
  \label{eq:67}
  \tilde {T^\mu}_\nu \xi^\nu =\delta^\mu_0 \frac{k}{\pi}{\rm
    Tr}\big[-\half (\lambda^{-1}\lambda')^2 T - \lambda'\lambda^{-1}
  \alpha' Y + u(\lambda^{-1}\lambda')(\lambda^{-1}\lambda')' Y\big]. 
\end{equation}

The chiral WZW theory for flat space \eqref{eq:54a} can be understood
as a flat limit of the sum of chiral $\mathfrak{sl}(2,\mathbf R)$ WZW
theories described by $I_\pm$ in \eqref{eq:15}. In order to take the
limit in terms of a dimensionless parameter we replace in $I_\pm$ the
cosmological radius $l$ by $l^{\epsilon}=\epsilon^{-1} l$ and $G_\pm$
by $G_\pm^\epsilon$ involving an explicit $\epsilon$ dependence. If we
assume $G^\epsilon_\pm=\Lambda(\mathbb 1\pm \epsilon \frac{b_\pm}{l}
)+O(\epsilon^2)$ with $a=\half \Lambda(b_++b_-)\Lambda^{-1}$, we have
$\lim_{\epsilon\to 0} \frac{kl}{2\pi\epsilon}
(\Gamma[G^\epsilon_+]-\Gamma[G^\epsilon_-])=\frac{k}{2\pi}
\Gamma[\Lambda,a]$, while for the two dimensional term one gets
$\frac{k}{2\pi} {\rm Tr}(\Lambda ' \Lambda^{-1} \dot a + \dot \Lambda
\Lambda^{-1} a' )$ as $\epsilon \to 0$. Summing up both contributions
and using \eqref{eq:54} gives the result.

\section{Reduction to Liouville}
\label{sec:Liouville}

\subsection{AdS case}
\label{sec:ads-case}

Let us discuss the reduction at the level of the chiral WZW actions.

In Fefferman-Graham gauge, one can read from equations \eqref{eq:22}
that $(g^{-1}_\pm\d_\pm g_\pm)^\mp=1$ where the superscript denotes
the component along the Lie algebra element $E_-$ respectively $E_+$.
Using in addition $\d_\pm g_\mp=0$, this implies the conditions
$(g^{-1}_+g_+^{\prime})^-=1$ while $(g^{-1}_- g_-^{\prime})^+=1$,
which correspond to fixing some of the chiral conserved current
components in \eqref{eq:43}. 

In terms of the parametrization
\begin{equation}
  \label{eq:82}
  g_\pm=e^{\sigma_\pm E_\mp} e^{\mp \half \varphi_\pm H} e^{\tau_\pm
    E_\pm}, 
\end{equation}
actions \eqref{eq:15} read
\begin{equation}
  \label{eq:84}
  I_\pm[g_\pm]=\pm\frac{k}{4\pi}\int dud\phi \big[
  \varphi_\pm^{\prime}\d_\mp\varphi_\pm
-4  e^{-\varphi_\pm}\sigma_\pm^{\prime}\d_\mp \tau_\pm\big],
\end{equation}
while the reduction conditions become
$e^{-\varphi_\pm}\sigma_{\pm}^{\prime}=\pm1$.  Up to boundary terms,
the reduced actions are the ones for chiral bosons,
\begin{equation}
  \label{eq:86}
I^R_\pm=\pm\frac{k}{4 \pi}\int du d\phi\big(
\varphi_\pm^{\prime}\d_\mp\varphi_\pm \big). 
\end{equation}

In the BMS gauge, we use the parametrization 
\begin{equation}
g_\pm=e^{\pm \frac{\sigma_\pm}{\sqrt{2}} E_-} e^{-\half \varphi_\pm H}
e^{\pm \sqrt{2}\tau_\pm E_+}\label{eq:87},
\end{equation}
which, when inserted into actions \eqref{eq:15} gives the same actions
\eqref{eq:84}. The reduction conditions $(f^{-1}_\pm
f_\pm^{\prime})^-= \frac{1}{\sqrt 2}$ then become again
$e^{-\varphi_\pm}\sigma_{\pm}^{\prime}=\pm1$ and give rise to the same
reduced actions \eqref{eq:86} as in the previous gauge.

As shown in Section~\bref{sec:group-elements}, the sum of the chiral
boson actions
\begin{equation}
\label{chiralbosons}
I^R_++I^R_-=\frac{k l}{8\pi} \int du 
 d\phi \left(\dot{\varphi}_+\varphi_+'-\frac{1}{l}(\varphi_+')^2
 -\dot{\varphi}_-\varphi_-'-\frac{1}{l}(\varphi_-')^2\right),
\end{equation}
can be written in Liouville form \eqref{eq:111} by using
transformation \eqref{eq:29}.

The reduction of the chiral models to Liouville can also be discussed
in terms of the modified Sugawara construction. By allowed
redefinitions of the currents associated to conformal transformations
as discussed in the previous section, equivalent representatives for
the time components can be chosen as $\tilde T^\pm_{\pm\pm}=T^\pm_{\pm\pm}
+\mu^a_\pm \d_\phi I^\pm_a$. Note that in this case, the
spatial parts of the currents have to be modified accordingly. On the
surface defined by the reduction constraints, the only representatives
that commute with the first class reduction constraints
$I_0^+=-\frac{kl}{4\pi} \sqrt 2$, $I_1^-=\frac{kl}{4\pi}\sqrt 2$ (FG
gauge), respectively $I_0^\pm=\mp\frac{kl}{4\pi}$ (BMS gauge) are
$\tilde T^\pm_{\pm\pm}\approx T^\pm_{\pm\pm} \pm l \d_\phi I^\pm_2$. Being
first class, these are observables of the reduced theory and their
Dirac brackets in Liouville theory coincide, on the constraint
surface, with their brackets in the chiral WZW models, which are
explicitly given by
\begin{equation}
  \label{eq:81}
  \{\tilde T^\pm_{\pm\pm}(\phi),\tilde T^\pm_{\pm\pm}(\phi')\}^*=\pm l
  (\tilde T^\pm_{\pm\pm}(\phi)+\tilde T^\pm_{\pm\pm} (\phi'))\d_\phi
  \delta(\phi-\phi')\mp\frac{k l^3}{4\pi} \d^3_\phi\delta(\phi-\phi').
\end{equation}
In terms of modes $L^\pm_m= \frac{1}{l} \int^{2\pi}_0 d\phi e^{\pm
  im\phi} \tilde T^\pm_{\pm\pm}$, this gives the standard Dirac
bracket algebra
\begin{equation}
\begin{split}
& i\{L^\pm_n,L^\pm_n\}^*=(m-n)L^\pm_{m+n}+\frac{c}{12}
m^3\delta^0_{m+n},\quad c=6kl=\frac{3l}{2G},
\\ & i\{L^\pm_n,L^\mp_n\}^*=0. 
\end{split}
\label{eq:33}
\end{equation}

\subsection{Flat case}
\label{sec:flat-case-1}

From equation \eqref{eq:98} and \eqref{eq:37}, the reduction
conditions are
\begin{equation}
(\lambda^{-1}\lambda')^-\approx\frac{1}{\sqrt 2}=\frac{2 \pi}{k}
P^-,\quad J^-\approx  0\Rightarrow (\lambda^{-1}\alpha'\lambda)^-\approx 0 . \label{eq:99}
\end{equation}

In terms of the parametrization,
  \begin{eqnarray}
\lambda=e^{\frac{1}{\sqrt{2}}\sigma E_{-}}e^{-\frac{1}{2}\varphi H}
e^{\sqrt{2} \tau E_{+}}, 
\quad \alpha= \frac{\eta}{\sqrt{2}}E_{+}+\frac{\theta}{2}H+\frac{\zeta}{\sqrt{2}}E_{-},
 \end{eqnarray}
the flat chiral WZW action \eqref{eq:54a} reads
  \begin{multline}
 I[\varphi,\sigma,\tau,\eta,\theta,\zeta]=\frac{k}{2\pi} \int du d\phi
 \Big[-(\theta'+\sigma \eta')\dot{\varphi}+
\eta'\dot{\sigma}-\\-(\eta'\sigma^2+2\theta'\sigma-2\zeta')e^{-\varphi}\dot{\tau} 
 -\frac{1}{2}\varphi'^2-2\sigma'\tau'e^{-\varphi}\Big],
 \end{multline}
while the reduction conditions become
\begin{equation}
  \label{eq:102}
  \sigma'e^{-\varphi}\approx 1,\quad \eta'\sigma^2+
2\theta'\sigma-2\zeta'\approx 0. 
\end{equation}

Using integration by parts and neglecting all boundary terms, the
reduced action can be written as
 \begin{equation}
 I=\frac{k}{4\pi}\int du d\phi
 \Big[\xi'\dot{\varphi}
 -\varphi'^2\Big],
 \end{equation}
 where $\xi=-2(\theta + \sigma \eta)$.  This is the centrally extended
 BMS$_3$ invariant action in the form \eqref{eq:42}. Again, as shown
 in Section~\bref{sec:group-elements}, it is related to the
 Liouville-like form \eqref{eq:113} through the transformation
 \eqref{eq:38}.

 The analog of the modified Sugawara construction for the flat case is
 as follows. The time-components of the currents associated to ${\rm
   BMS}_3$ transformations may be redefined as $\tilde \cH=\cH
 +\mu^a\d_\phi P_a+\nu^a\d_\phi J_a$ and $\tilde \cP=\cP+\rho^a
 \d_\phi P_a+\sigma^a\d_\phi J_a$. Note that in this case, the spatial
 parts of the currents do not need to be modified since $\d_0 P\approx
 0 \approx \d_0 J$. On the surface defined by the reduction
 constraints, the only representatives that commute with the first
 class reduction constraints $J_0=0$, $P_0=\frac{k}{2\pi}$ are $\tilde
 \cH \approx \cH+\d_\phi P_2$ and $\tilde \cP\approx \cP-\d_\phi J_2$.
 Being first class, these are observables of the reduced theory and
 their Dirac brackets in BMS Liouville theory coincide, on the
 constraint surface, with their brackets in the chiral
 $\mathfrak{iso}(2,1)$ WZW like model, which are given by
\begin{equation}
  \label{eq:62a}
  \begin{split}
    & \{\tilde \cH(\phi),\tilde \cH(\phi')\}^*=0,\\
    & \{\tilde \cH(\phi),\tilde \cP(\phi')\}^*=(\tilde
    \cH(\phi)+\tilde \cH(\phi'))\d_\phi\delta(\phi-\phi')-\frac{k}{2\pi}
    \d_\phi^3\delta(\phi-\phi')
    ,\\
    & \{\tilde \cP(\phi),\tilde \cP(\phi')\}^*=(\tilde
    \cP(\phi)+\tilde \cP(\phi'))\d_\phi\delta(\phi-\phi').
  \end{split}
\end{equation}
In terms of modes, $P_m=\int^{2\pi}_0 d\phi e^{im\phi} \tilde \cH$,
$J_m=\int^{2\pi}_0 d\phi e^{im\phi} \tilde \cP$, this gives the
centrally extended BMS3 algebra, 
\begin{equation}
  \label{eq:116}
  \begin{split}
    & i\{P_m,P_n\}^*=0,\\
    & i\{J_m,P_n\}^*=(m-n) P_{m+n}+\frac{c_2}{12}m^3\delta^0_{m+n}
    ,\quad c_2=12 k=\frac{3}{G}, \\
    & i\{J_m,J_n\}^*=(m-n)
    J_{m+n}+\frac{c_1}{12}m^3\delta^0_{m+n},\quad c_1=0. 
  \end{split}
\end{equation}

\section{A comment on zero modes}
\label{sec:comment-zero-modes}

As already stressed in \cite{Henneaux:1999ib}, the change of variables
\eqref{eq:36} is not well-defined in the zero mode sector. As a
consequence, the equivalence of the sum of the two chiral models with
the non-chiral theory does not hold in this sector. The same applies
to the transformation \eqref{eq:29} used in order to relate the sum of
two chiral bosons with Liouville theory, and also to the
transformation \eqref{eq:38} that relates a free chiral boson like
action to BMS Liouville theory.

It then follows that asymptotically AdS or flat gravity is, strictly
speaking, not equivalent to (BMS) Liouville theory, but rather to the
respective chiral models.

\section*{Acknowledgements}
\label{sec:acknowledgements}

\addcontentsline{toc}{section}{Acknowledgments}

The authors thank A.~Gomberoff for collaboration during the early
stages of this work and C.~Troessaert for useful discussions. This work is
supported in part by the Fund for Scientific Research-FNRS (Belgium),
by IISN-Belgium, by ``Communaut\'e fran\c caise de Belgique - Actions
de Recherche Concert\'ees'' and by Fondecyt Projects No.~1085322 and
No.~1090753. H.G.~thanks Conicyt for financial support.

\appendix

\section{Chern-Simons formulation of gravity}
\label{sec:chern-simons-form}

Let $A=-1,0,1,2$ and $a=0,1,2$ and consider the flat metric
$\eta_{AB}={\rm diag}(-1-1,1,1)$. In
terms of the (anti-hermitian) generators $P_a=\frac{1}{l} J_{-1
  a},J_{ab}$, the $\mathfrak{so}(2,2)$ algebra reads
\begin{equation}
  \label{eq:4}
\begin{split}
  [J_{ab},P_c]=\eta_{bc}P_a-\eta_{ac} P_b,\quad &
  [J_{ab},J_{cd}]=\eta_{bc} J_{ad}-\eta_{ac} J_{bd}-\eta_{bd}
  J_{ac}+\eta_{ad} J_{bc},\\ & [P_a,P_b]=\frac{1}{l^2} J_{ab}\,.
  \end{split}
\end{equation}
The three dimensional Poincar\'e algebra $\mathfrak{iso}(2,1)$is
obtained by keeping the generators $P_a,J_{ab}$ fixed and taking the
limit of $l$ to infinity.

Let $\epsilon_{012}=1$ and take $\eta_{ab}={\rm diag}(-1,1,1)$ and its
inverse to lower and raise tangent space indices $a,b,c,\dots$.  In
terms of $J^a=-\half \epsilon^{abc} J_{bc} \iff J_{ab}=\epsilon_{abc}
J^c$, the algebra reads
\begin{equation}
[J_a,J_b]=\epsilon_{abc} J^c,\quad [J_a,P_b]=\epsilon_{abc} P^c,\quad 
[P_a,P_b]=\frac{1}{l^2}\epsilon_{abc} J^c. 
\end{equation}

When neglecting boundary terms, the gravitational action in terms of
dreibeins $e^a=e^a_\mu dx^\mu$ and spin connection $\omega=\half
\omega^{ab}_\mu J_{ab} dx^\mu=\omega^a_\mu J_adx^\mu$ can be written
as 
\begin{equation}
  \label{eq:5}
  S[e,\omega  ]=\frac{1}{16\pi G}\int d^3 x\, e(
  e^\mu_ae^\nu_b R^{ab}_{\mu\nu}-2\Lambda)=-\frac{1}{8\pi G}\int (
  e_a R^a-\frac{\Lambda}{6} \epsilon_{abc} e^ae^b e^c),
\end{equation}
with $ \frac{1}{2} R^{ab} J_{ab} =d\omega +\omega^2 =R^a J_a$. We
always omit the wedge product and have chosen the orientation for the
integration of 3-forms according to $d^3x=drdud\phi$ so that the
boundary Wess-Zumino-Witten actions come with the standard sign. The
latter action is equivalent to the Chern-Simons action
\begin{equation}
  \label{eq:6}
  S[A]=-\frac{k}{4\pi}\int \langle A, dA+\frac{2}{3} A^2
  \rangle,
\end{equation}
where $A=\omega^a J_a+e^a P_a$, $\langle J_a, P_b\rangle =\eta_{ab}$,
$\langle J_a, J_b\rangle =0=\langle P_a, P_b\rangle$ and 
\begin{equation}
k=\frac{1}{4 G},\quad \Lambda=-\frac{1}{l^2}.\label{eq:108}
\end{equation}

In order to adapt the problem to our gauge choice, we now use
light-cone coordinates in tangent space by introducing two null
vectors,
\begin{equation}
  \label{eq:2}
  e_a^{\mu} e_{\mu b}=\eta_{ab}, 
\quad \eta_{ab}=\begin{pmatrix} 0 &
    1 & 0\\
    1  & 0 & 0 \\
    0 & 0 & 1
\end{pmatrix}.
\end{equation}
In the fundamental representation of $\mathfrak{sl}(2,\mathbb R)$,
generators satisfying $ j_a j_b =\half
\epsilon_{abc}j^c+\frac{1}{4}\eta_{ab}\mathbb{1}$,
${\rm Tr}(j_a j_b)=\half \eta_{ab}$,
$[j_a,j_b]=\epsilon_{abc} j^c$ are given
by
\begin{equation}
\label{generators}
j_0 = \frac{1}{\sqrt 2} \begin{pmatrix} 0&1\\ 0& 0 \end{pmatrix},\:
j_1 =\frac{1}{\sqrt 2} \begin{pmatrix} 0 & 0\\ 1&0 \end{pmatrix},\:
j_2= \frac{1}{2} \begin{pmatrix} 1&0\\ 0&-1 \end{pmatrix}.
\end{equation}
In terms of $e=e^a_\mu dx^\mu j_a$ and
$\omega=\omega^a_\mu dx^\mu j_a$, with
$\omega^{ab}=\epsilon^{abc}\omega_c$, the
explicit form of the equations of motion, the zero curvature condition
$F\equiv dA+A^2= 0$ is \eqref{eq:13}.

The chiral decomposition $J^\pm_a=\half (J_a\pm l P_a)$, $A^\pm=A^{\pm
  a} J^\pm _a$, $A^{a\pm}=\omega^a\pm \frac{1}{l} e^a$ disentangles
the algebra in terms of $\mathfrak{so}(2,1)\oplus \mathfrak{so}(2,1)$
and allows one to write the gravitational action \eqref{eq:5} as the
difference of two Chern-Simons terms,
\begin{equation}
  \label{eq:7}
  S[A^+,A^-]=-\frac{l k}{8\pi}\int \eta_{ab} \Big(
  A^{+a}\big[dA^{+}+\frac{2}{3} (A^{+})^2\big]^b-
  A^{-a}\big[dA^{-}+\frac{2}{3} (A^{-})^2\big]^b\Big). 
\end{equation}
It is only well-defined for non-zero cosmological constant, while all
previous considerations have a straightforward flat space limit $l\to
\infty$ for which $\mathfrak{so}(2,2)$ contracts to
$\mathfrak{iso}(2,1)$. 
Action \eqref{eq:7} can be written in matrix form as
\begin{equation}
S[A^+,A^-]=\frac{l}{2}(S_{CS}[A^+]-S_{CS}[A^-]),\
S_{CS}[A]=-\frac{k}{2\pi}\int {\rm Tr} (AdA+\frac{2}{3} A^3).\label{eq:107}
\end{equation}
We will use coordinates $r,u,\phi$ and, in the ${\rm
  AdS}$ case, $x^\pm=\frac{u}{l}\pm\phi$ with
$2\d_{\pm}=l\d_{u}\pm\d_{\phi}$.  Note that the redefinition
$l_{-1}=\sqrt 2 j_0$, $l_1=-\sqrt 2 j_1$, $l_0= j_2$ gives
$[l_m,l_n]=(m-n) l_{m+n}$ for $m,n=-1,0,1$ and $E_+=\sqrt 2 j_0$,
$E_-=\sqrt 2 j_1$, $H=2 j_2$ gives $[E_+,E_-]=H$, $[H,E_{+}]=2E_{+}$,
$[H,E_{-}]=-2E_{-}$ and
\begin{equation}
  \label{eq:58}
  \begin{split}
 &   e^{[x E_+,\cdot]}E_-=E_-+x H-x^2 E_+,\quad e^{[x E_+,\cdot]}H=H-2x
    E_+,\\
& e^{[y E_-,\cdot]}E_+=E_+-y H-y^2 E_-,\quad e^{[y E_-,\cdot]}H=H+2y
    E_-,\\
& e^{[\half z H,\cdot]}E_+=e^z E_+,\quad e^{[\half z H,\cdot]}E_-=e^{-z} E_-.
  \end{split}
\end{equation}

\section{Wess-Zumino-Witten theories}
\label{sec:wess-zumino-witten}

\subsection{Generalites}

For the two dimensional Wess-Zumino-Witten theories, we will use
coordinates $u,\phi$ with $\eta_{\mu\nu}={\rm diag}(-1,l^2)$ and
$\epsilon^{\mu\nu}$ determined by $\epsilon^{01}=1$. In the light-cone
basis, we have $\eta^{+-}=-\frac{2}{l^2}=\eta^{-+}$ and
$\eta^{++}=0=\eta^{--}$, while $\epsilon^{+-}=-\frac{2}{l}$.

For factorized group elements $G=g(\phi,u)h(r,u)$ satisfying $\dot
h(\infty,u)=0$, we have $\d_\pm G G^{-1}= - (\d_\pm g h + \frac{l}{2}g
\dot{h})h^{-1}g^{-1} =- \d_\pm g g^{-1}$. With only the boundary at
$r=\infty$ one has for the Wess-Zumino-Witten term $\Gamma[G]$ defined
in \eqref{eq:46}
\begin{equation}
  \label{eq:69}
  \delta\Gamma[G]=\half\int dud\phi{\rm Tr} (\epsilon^{\mu\nu} 
  \delta g g^{-1}\d_\mu g g^{-1}\d_\nu gg^{-1}),
\end{equation}
with $\epsilon^{01}=1$ and where we have assumed that $\delta G=\delta
g h$. Furthermore
\begin{multline}
  \label{eq:68}
  \Gamma[G^{-1}_1G_2]=\Gamma[G_1^{-1}]+\Gamma[G_2]+\\+ \frac{1}{l}\int du d\phi
 {\rm Tr} [\d_- g_1 g^{-1}_1\d_+ g_2 g_2^{-1}- \d_+ g_1g^{-1}_1\d_- g_2 g_2^{-1}], 
\end{multline}
while the WZW action
defined in \eqref{eq:34} satisfies the Polyakov-Wiegmann identities
\begin{equation}
    I[g^{-1}h]=I[g^{-1}]+I[h]-\frac{k}{\pi} \int dud\phi {\rm Tr} \Big[\d_-G G^{-1}
    \d_+ H H^{-1}\Big]^{r\to \infty},
\label{eq:61a}
\end{equation}

Under the same assumptions, for the chiral Wess-Zumino-Witten theories
defined in \eqref{eq:15} we get instead,
\begin{multline}
  \label{eq:41}
  I_\pm[g_\pm^{-1} h_\pm]=I_\pm [g_\pm^{-1}]+I_\pm[h_\pm]\mp \\
  \mp\frac{k}{\pi}\int dud\phi {\rm Tr}\Big[ (\d_+ G_\pm
  G^{-1}_\pm-\d_-G_\pm G^{-1}_\pm)\d_\mp H_\pm H^{-1}_\pm\Big]^{r\to
    \infty}.
\end{multline}

\subsection{Group elements and Poisson brackets}

Introduce local coordinates $\zeta^a$ on the group manifold. We
have 
\begin{equation}
\begin{split}
& g^{-1}dg=\theta^a
  j_a,\ \theta^a={M^a}_b(\zeta)
  d\zeta^b,\ d\theta^a=-\half \epsilon^a_{bc}
  \theta^b\theta^c,\\
& dg g^{-1}= \kappa^a j_a,\ \kappa^a={{N}^a}_b(\zeta)
d\zeta^b,\ d\kappa^a=\half
\epsilon^a_{bc}\kappa^b\kappa^c,
\end{split}
\end{equation}
where
\begin{equation}
\begin{split}
& {N^a}_b={K^a}_c
{M^c}_b,\ {K^a}_c=2{\rm Tr}(j^a g j_c
g^{-1} ),\ {K^{-1ac}}=2{\rm Tr}(j^a g^{-1} j^c g
)=K^{ca},\\ & {\epsilon_{cd}}^e
{(K^{-1})^c}_a
{(K^{-1})^d}_b={\epsilon_{ab}}^f{(K^{-1})^e}_f,
\\ & {\epsilon^{ab}}_e
{(K^{-1})^c}_a
{(K^{-1})^d}_b={\epsilon^{cd}}_f
{(K^{-1})^f}_e.
\end{split}
\end{equation}
Locally, 
\begin{equation}
\Gamma[G]=\frac{1}{2}\int dud\phi\
\epsilon^{\mu\nu}B_{ab}\d_\mu\zeta^a\d_\nu
\zeta^b,\quad B_{ab}(\zeta)=-B_{ba}(\zeta).
\end{equation} 
Let $H_{abc}=\d_a B_{bc}+({\rm
  cyclic}\ a,b,c)$. From the variation of $\Gamma$, one
has
\begin{equation}
  \label{eq:78}
\begin{split}
  H_{abc}=\half {\rm Tr}\Big(g^{-1}\ddl{g}{\zeta^a}
[g^{-1}\ddl{g}{\zeta^b},g^{-1}\ddl{g}{\zeta^c}]\Big)=\frac{1}{4}
{M^g}_a{M^e}_b{M^f}_c\epsilon_{gef}\\=
\half {\rm Tr}\Big(\ddl{g}{\zeta^a}g^{-1}
[\ddl{g}{\zeta^b}g^{-1},\ddl{g}{\zeta^c}g^{-1}]\Big)=\frac{1}{4}
{N^g}_a{N^e}_b{N^f}_c\epsilon_{gef}.
\end{split}
\end{equation}
Consider the canonical momenta $\eta=\eta_a j^a$,
$\{\zeta^a(\phi),\eta_b(\phi')\}=\delta^a_b\delta(\phi-\phi')$
and define $\pi= -\eta_b
{(M^{-1})^b}_a j^a$, the Poisson brackets are
\begin{equation}
  \label{eq:72c}
\begin{split}
  & \{g(\phi),\pi_a(\phi')\}=- g(\phi) j_a
  \delta(\phi-\phi^\prime),\\
  &
  \{\pi_a(\phi),\pi_b(\phi')\}={\epsilon_{ab}}^c
  \pi_c(\phi)\delta(\phi-\phi^\prime),\\
  &
  \{(g^{-1}g')_a(\phi),\pi_b(\phi')\}={\epsilon_{ab}}^c(g^{-1}
  g')_c(\phi)\delta(\phi-\phi^\prime)-\eta_{ab} \d_\phi
  \delta(\phi-\phi^\prime).
\end{split}
\end{equation}
Similarly, with $\rho=\eta_b{(N^{-1})^b}_a
j^a$, 
\begin{equation}
  \label{eq:72a}
\begin{split}
&  \{g(\phi),\rho_a(\phi')\}=  j_a g(\phi)
  \delta(\phi-\phi^\prime),\\
& \{\rho_a(\phi),\rho_b(\phi')\}={\epsilon_{ab}}^c
\rho_c(\phi)\delta(\phi-\phi^\prime),\\
& \{(g' g^{-1})_a(\phi),\rho_b(\phi')\}={\epsilon_{ab}}^c(
g'g^{-1})_c(\phi)\delta(\phi-\phi^\prime)+\eta_{ab}
\d_\phi \delta(\phi-\phi^\prime),\\
& \{\pi_a(\phi),\rho_b(\phi')\}=0.
\end{split}
\end{equation}

\subsection{Current algebra of the non-chiral WZW theory}
\label{sec:current-algebra-non}

In local coordinates, the Lagrangian density for the non-chiral WZW
action is given by 
\begin{equation}
  \label{eq:26}
  \frac{16 \pi}{kl^2} \cL= M_{a  b}
  M^{a}\,_{e}(\dot{\zeta}^{b}\dot{\zeta}^{e}
-\frac{1}{l^2}\zeta^{b \prime} \zeta^{e \prime})-
\frac{8}{l}B_{a b}\dot{\zeta}^{a}\zeta^{b \prime}.
\end{equation}
The relation between canonical momenta $\eta_g$ and velocities is
\begin{equation}
\eta_{c}\approx \frac{\d \mathcal{L}}{\d (\d_{0}
  \zeta^{c})}=\frac{kl^2}{8\pi}(M^{a}\,_{c}
(g^{-1}\dot{g})_{a}-\frac{4}{l}B_{c b}\zeta^{b
  \prime}).\label{eq:55}
\end{equation}
Defining $v_{c}=\eta_{c}+\frac{kl}{2\pi}B_{c
  b}\zeta^{b \prime }$, we have
\begin{equation}
  \label{eq:current01}
 \{v_a(\phi),v_b(\phi')\}=-\frac{kl}{2\pi} 
H_{abc}\zeta^{c\prime}(\phi)
\delta(\phi-\phi^\prime).
\end{equation}
The Hamiltonian is
\begin{equation}
  \label{eq:104}
  H=\frac{4\pi}{kl^2}  {(M^{-1})^a}_b
  {(M^{-1})^{cb}} v_a v_c+\frac{k}{16\pi}
  M_{ba}{M^b}_c
  \zeta^{a\prime}\zeta^{c\prime}. 
\end{equation}
In terms of the improved momenta $\pi^B=
-v_b{(M^{-1})^b}_a j^a \approx -\frac{kl^2}{8 \pi}
g^{-1}\dot{g}$, the Poisson brackets are the same
as in \eqref{eq:72c} with $\pi$ replaced by $\pi^B$, 
except for
\begin{equation}
  \label{eq:current02}
\{\pi^B_a(\phi),\pi^B_b(\phi')\}={\epsilon_{ab}}^c
(\pi^B_c- \frac{k
  l}{8\pi}(g^{-1} g')_c)(\phi)\delta(\phi-\phi^\prime),
\end{equation}
while the first order action principle can be written as 
\begin{equation}
  \label{eq:115}
  I_H=-\int dud\phi\, {\rm Tr}\, \Big[2\pi^B g^{-1}\dot g +\frac{k}{8\pi}
  (g^{-1}g')^2 +\frac{8\pi}{kl^2}(\pi^B)^2\Big]-\frac{kl}{2\pi}\Gamma[G]. 
\end{equation}
The current components $I_-$ of \eqref{eq:101} are given on-shell by
$I_-\approx \pi^B+\frac{kl}{8\pi} g^{-1}g'$ so that $\int d\phi^\prime
2{\rm Tr} [I_-\theta_-]$ is the canonical generators of the symmetry
transformation $\delta g=-g\theta_-$. Evaluating the Poisson brackets
then leads to \eqref{eq:74b} for $I_-$.

Defining $\rho^B=v_b{(N^{-1})^b}_a j^a$. Since in
\eqref{eq:26} and \eqref{eq:55} one can replace ${M^a}_b$ by
${N^a}_b$, we have $\rho^B\approx \frac{kl^2}{8\pi} \dot g
g^{-1}$. The Poisson bracket are as in \eqref{eq:72a} except for
\begin{equation}
 \{\rho^B_a(\phi),\rho^B_b(\phi')\}={\epsilon_{ab}}^c
(\rho^B_c- \frac{k
  l}{8\pi}(g' g^{-1})_c)(\phi)\delta(\phi-\phi^\prime).
\end{equation}
On shell $I_+$ of \eqref{eq:101} is now given by $I_+\approx
\rho^B+\frac{kl}{8\pi} g'g^{-1}$ so that $\int d\phi^\prime 2{\rm Tr}
[I_+\theta_+]$ is the canonical generators of the symmetry
transformation $\delta g=\theta_+ g$. Evaluating the Poisson brackets
then leads to \eqref{eq:74b} for $I_+$. In the same way, one then
establishes that the left and right current components have vanishing
Poisson brackets.

\subsection{Current algebra of the chiral models}
\label{sec:curr-algebra-chir}

Since a different parametrization for the right and left group
elements will be useful, we will use $\zeta^a_\pm$ in what
follows. The local Lagrangian densities are
\begin{multline}
  \label{eq:75}
  \frac{8\pi}{k l}\cL^\pm= \pm (g^{-1}_\pm g'_\pm)_a (g^{-1}_\pm
  \dot g_\pm)^a -\frac{1}{l} (g^{-1}_\pm g'_\pm)_a
  (g^{-1}_\pm g'_\pm)^a \pm 4
  B^\pm_{ab}\dot\zeta_\pm^a \zeta_\pm^{b\prime}.
\end{multline}
The canonical momenta are related to the velocities through
\begin{equation}
\eta^\pm_a\approx \ddl{\cL^\pm}{\d_0\zeta_\pm^a}\approx \pm \frac{k
  l}{8\pi}\Big( (g^{-1}_\pm g'_\pm)_b {M^b_\pm}_\alpha+4
B^\pm_{ab}\zeta_\pm
^{b\prime}\Big)\label{eq:90}.
\end{equation}
Defining
\begin{equation}
  \label{eq:76}
  v^\pm_a= \eta^\pm_a \mp \frac{k l}{2\pi} 
B^\pm_{ab}\zeta^{b\prime}_\pm,
\end{equation}
we now have
\begin{equation}
  \label{eq:77}
 \{v^\pm_a(\phi),v^\pm_b(\phi')\}=\pm \frac{k l}{2\pi} 
H^\pm_{abc}\zeta^{c\prime}_\pm(\phi)
\delta(\phi-\phi^\prime),
\end{equation}
with Poisson brackets of variables associated to different chiral
copies all commuting. In terms of the improved momenta $\pi^{B\pm}= -v^\pm_b
{(M^{-1})^b_\pm}_a j^a$, the Poisson brackets are as in
\eqref{eq:72c} for each copy except for 
\begin{equation}
  \label{eq:72}
\{\pi^{B\pm}_a(\phi),\pi^{B\pm}_b(\phi')\}={\epsilon_{ab}}^c
(\pi^{B\pm}_c\pm \frac{k
  l}{8\pi}(g^{-1}_\pm g'_\pm)_c)(\phi)\delta(\phi-\phi^\prime). 
\end{equation}
The primary constraints can be written as 
\begin{equation}
  \label{eq:71}
  \phi_\pm=\pi^B_\pm\pm \frac{k l}{8\pi} g^{-1}_\pm g_\pm'\approx 0.
\end{equation}
Consider then 
\begin{equation}
I_\pm=\pi^B_\pm\mp\frac{k l}{8\pi} g^{-1}_\pm
g'_\pm\label{eq:103}.
\end{equation}
They agree on the constraint surface with the time components of the
conserved currents \eqref{eq:43a}, $I_\pm\approx \mp \frac{k l}{4\pi}
g_\pm^{-1}g'_\pm$. Furthermore, $\int d\phi'\, 2{\rm Tr}[I_\pm
\theta_\pm]$ are the canonical generators of the symmetry
transformations $\delta_{\theta_\pm} g_{\pm}=-g_{\pm}\theta_\pm$. The
components of $I_\pm$ satisfy the current algebra \eqref{eq:74b} in
the standard Poisson bracket and have weakly vanishing Poisson
brackets with the constraints, i.e., they are first class,
\begin{equation}
  \label{eq:73}
  \{I^\pm_a(\phi),\phi^\pm_b(\phi^\prime)\}={\epsilon_{ab}}^c 
  \phi^\pm_c(\phi)
  \delta(\phi-\phi^\prime).
\end{equation}
This proves the result on the Dirac brackets of the chiral currents.

Defining $\rho^B_\pm=v^\pm_b{(N^{-1}_\pm)^b}_a
j^a$, the Poisson bracket are as in \eqref{eq:72a} except for 
\begin{equation}
 \{\rho^{B\pm}_a(\phi),\rho^{B\pm}_b(\phi')\}={\epsilon_{ab}}^c
(\rho^{B\pm}_c\pm \frac{k
  l}{8\pi}(g'_\pm g^{-1}_\pm)_c)(\phi)\delta(\phi-\phi^\prime). 
\end{equation}
Since $p^\pm_a\approx\ddl{\cL^\pm}{\d_0\zeta_\pm^a}= \pm
\frac{k l}{8\pi}\Big( ( g'_\pm g^{-1}_\pm)_b
{N^b_\pm}_a+4 B^\pm_{ab}\zeta_\pm
^{b\prime}\Big)$, the primary constraints can be written as
\begin{equation}
  \label{eq:70}
  \psi_\pm= \rho^B_\pm\mp \frac{k l}{8\pi} g'_\pm g^{-1}_\pm
 \approx 0.
\end{equation}
They satisfy the current algebra
\begin{equation}
  \label{eq:74a}
  \{\psi^\pm_a(\phi),\psi^\pm_b(\phi^\prime)\}=
  {\epsilon_{ab}}^c \psi^\pm_c(\phi)
\delta(\phi-\phi^\prime)\mp
\frac{k l}{4\pi}\eta_{ab}\d_\phi\delta(\phi-\phi^\prime). 
\end{equation}
In terms of this representation of the constraints, 
\begin{equation}
  \label{eq:79}
   \{I^\pm_a(\phi),\psi^\pm_b(\phi^\prime)\}=0.
\end{equation}
It follows from \eqref{eq:74a} that the zero modes
$\Psi^\pm_a=\int_{0}^{2\pi} d\phi\ \psi^\pm_a$ are first
class constraints that generate the arbitrary function of $u$ in the
general solution to the equations of motion, while all other modes are
second class constraints.

The chiral Hamiltonians are 
\begin{equation}
  \label{eq:48}
\begin{split}
  H^\pm  & =\frac{k}{4\pi} \int d\phi\, {\rm Tr}\, [ g'_\pm g^{-1}_\pm
   g'_\pm g^{-1}_\pm]+\int d\phi\,  2{\rm Tr}\, [u_\pm \psi_\pm]\\
& = \frac{k}{4\pi} \int d\phi\, {\rm Tr}\, [ g^{-1}_\pm g'_\pm 
   g^{-1}_\pm g'_\pm ]+\int d\phi\,  2{\rm Tr}\, [v_\pm \phi_\pm],
 \end{split}
\end{equation} 
where $u_\pm=u^a_\pm j_a, v_\pm=v^a_\pm j_a$
contain the Lagrange multipliers. Taking the Poisson bracket with the
primary constraints shows that there are no secondary ones.

\subsection{Current algebra of the flat model}
\label{sec:current-algebra-flat}

Locally,
\begin{equation}
  \label{eq:88}
  \frac{2\pi}{k} \cL= \dot \zeta^b
  {N^a}_b\alpha'_a-\half {N^a}_b
  N_{ac}\zeta^{b\prime}\zeta^{c\prime},
\end{equation}
and so, if $\eta_a,\omega_a$ are the momenta conjugate to
$\zeta^a,\alpha^a$, the primary constraints are 
\begin{equation}
  \label{eq:89}
  \eta_a\approx \frac{k}{2\pi} {N^b}_a\alpha'_b,\quad
  \omega_a\approx 0. 
\end{equation}
The primary constraints can be written as 
\begin{equation}
  \label{eq:91}
  \psi=\rho-\frac{k}{2\pi} \alpha'\approx 0,\quad \omega\approx 0. 
\end{equation}
Up to zero modes, they are second class since their algebra is 
\begin{equation}
  \label{eq:92}
\begin{split}
&  \{\psi_a(\phi),\psi_b(\phi')\}=
{\epsilon_{ab}}^{c}\rho_c\delta(\phi-\phi'),\\
& \{\psi_a(\phi),\omega_b(\phi')\}=-\frac{k}{2\pi}
\eta_{ab}\d_\phi\delta(\phi-\phi'),\\
& \{\omega_a(\phi),\omega_b(\phi')\}=0. 
\end{split}
\end{equation}
Consider 
\begin{equation}
\begin{split}
& P=\lambda^{-1} \omega \lambda+\frac{k}{2\pi} \lambda^{-1}\lambda',\\
& J= -\lambda^{-1}\rho\lambda+u P'=\pi+u P'.
\label{eq:95}
\end{split}
\end{equation}
On the constraint surface, they agree with the time components of the
Noether currents. Furthermore, $\int d\phi' 2 {\rm Tr} [P \sigma]$,
$\int d\phi' 2 {\rm Tr} [J \theta]$ are the canonical generators of
the infinitesimal symmetry transformations, $\delta_\sigma \lambda=0$,
$\delta_\sigma\alpha=\lambda \sigma \lambda^{-1}$ and $\delta_\theta
\lambda=-\lambda\theta$,
$\delta_\theta\alpha=-u\lambda\theta'\lambda^{-1}$.

They have weakly vanishing Poisson brackets with the constraints,
\begin{equation}
  \label{eq:93}
\begin{split}
&  \{P(\phi),\psi_b(\phi')\}=\lambda^{-1} [\omega
,j_b]\lambda\delta(\phi-\phi'), \quad 
\{P(\phi),\omega_b(\phi')\}=0,\\
&\{J(\phi),\psi_b(\phi')\}=u(\lambda^{-1}
[\omega,j_b]\lambda\delta(\phi-\phi'))',
\quad \{J(\phi),\omega_b(\phi')\}=0,
\end{split}
\end{equation}
and by direct computation, one finds that their Poisson brackets, and
thus also their Dirac brackets, form the $\mathfrak{iso}(2,1)$ current
algebra given in \eqref{eq:94}.

The Hamiltonian of the model is 
\begin{equation}
  \label{eq:60}
  H=\frac{k}{2\pi}\int d\phi {\rm Tr}[\lambda'\lambda^{-1} \lambda'\lambda^{-1} ]+\int d\phi
2 {\rm Tr}[u\psi +v \omega]. 
\end{equation}
Again, there are no secondary constraints. 


\def\cprime{$'$}
\providecommand{\href}[2]{#2}\begingroup\raggedright\endgroup

\end{document}